\documentclass[aps,a4paper,amsmath,amssymb,twocolumn,
floatfix,groupedaddress,showkeys ]{revtex4-1}
\usepackage{bm,amsmath,amssymb,amsfonts}
\usepackage[dvips]{graphicx}
\usepackage{color}
\usepackage{xcolor}
\usepackage{hyperref}

\hypersetup{
    pdfnewwindow=true,      
    colorlinks=true,       
    linkcolor=magenta,          
    citecolor=blue,        
    filecolor=blue,      
    urlcolor=blue        
}
\newcommand\bea{\begin{eqnarray}}
\newcommand\eea{\end{eqnarray}}
\newcommand\beq{\begin{equation}}
\newcommand\eeq{\end{equation}}

\begin{document}

\title{Aharonov-Bohm caging of an electron in a quantum fractal}
\author{Biplab Pal} 
\email[E-mail: ]{biplab@nagalanduniversity.ac.in}
\affiliation{Department of Physics, School of Sciences, 
Nagaland University, Lumami-798627, Nagaland, India}
\date{\today}
\begin{abstract}
Fractal geometries exhibit complex structures with scale invariance 
self-similar pattern over various length scales. An artificially designed 
quantum fractal geometry embedded in a uniform magnetic flux has been 
explored in this study. It has been found that due to quantum mechanical 
effect, such quantum fractal display an exotic electronic property which 
is reflected in its transport characteristics. Owing to this uniform 
magnetic flux piercing through each closed-loop building block of the 
fractal structure, an electron traversing through such a fractal geometry 
will pick up a nontrivial Aharonov-Bohm phase factor, which will influence 
its transport through the system. It is shown that, one can completely 
block the transmission of an electron in this fractal geometry by setting 
the value of the uniform magnetic flux to half flux quantum. This phenomenon 
of Aharonov-Bohm caging of an electron in this quantum fractal geometry has 
been supported by the computation of the energy spectrum, two-terminal 
transport and persistent current in its various generations. This result 
is very robust against disorder and could be useful in designing efficient 
quantum algorithms using a quantum fractal network.    
\end{abstract}
\keywords{Fractal; Aharonov-Bohm caging; Tight-binding model; 
Quantum transport; Persistent current}
\maketitle
%
\section{Introduction}
\label{sec:intro}
The lattice geometry and quantum mechanical effects play a pivotal role in 
controlling the quantum transport properties in low-dimensional quantum systems. 
The quantum transport property of a quantum system render a very significant role 
in processing the quantum information in the system as well as for building 
efficient quantum algorithms. The nature of the electronic states in the 
system, \textit {i.e.}, whether they are localized or extended, greatly 
influences the quantum transport properties of the system. For example, with 
any arbitrary amount of disorder in one- and two-dimensional quantum systems, 
all the single-particle states exhibit the well-known phenomena of Anderson 
localization~\cite{Anderson-prb-1958,Ramakrishnan-prl-1979}, 
where all the single-particle states are exponentially localized 
making the system to act as an insulator 
with zero transport. However, we can also have situations where one can have 
complete localization of an electronic state in absence of any disorder in 
the system. Such disorder-free localization of electronic states happens purely 
because of the underlying lattice topology~\cite{Sutherland-prb-1986} and 
the destructive quantum interferences of the wavefunction. This phenomenon 
has led to a very exciting area of research known as the flat-band 
localization~\cite{FB-review-Flach,FB-review-Vicencio}. 

Interestingly, one of the prototype sub-category examples of such flat-band 
localization for certain lattice geometries can emerge even in presence of 
an external magnetic flux in the system~\cite{Pal-prb-2018,
Pal-jpcm-2025,Bercioux-pra-2011,Vidal-prl-2000,Vidal-prl-1998}. 
The origin of this phenomenon lies in some fundamental aspects of quantum 
mechanics, where a charge particle moving around a closed-loop geometry 
pierced by a nonzero magnetic flux, picks up a geometric phase factor 
in its wavefunction known as the Aharonov-Bohm phase~\cite{AB-Phase}. For 
certain specific lattice geometries, such as one-dimensional (1D) diamond 
lattice~\cite{Vidal-prl-2000,Aharony-prb-2008,Aharony-physE-2010} 
and two-dimensional (2D) $\mathcal{T}_3$ 
lattice~\cite{Vidal-prl-1998}, we encounter a very special situation where 
the single-particle energy spectrum collapses to a few points for certain 
special value of the underlying magnetic flux equal to half flux quantum. 
This phenomenon is called as the Aharonov-Bohm (AB) caging, where all the 
single-particle states in the system are highly localized giving rise to 
zero transport. In recent times, this remarkable quantum effect has drawn 
a prevailing research attention and has been experimentally realized with 
the state-of-art ultrafast laser-writing technology in photonics using 
a synthetic magnetic flux~\cite{Sebabrata-prl-2018,Rodrigo-prl-2022} and 
also in optical lattice of ultracold atoms with tailored gauge 
fields~\cite{Longhi-prl-2022}. Lately, the idea of AB caging has also been 
extended for a variety of decorated lattice geometries in the context 
of topological insulator in lattice models~\cite{Amrita-jpcm-2021,
Dias-prr-2023,Kremer-natcomm-2020}. 
%
\begin{figure*}[ht]
\textbf{(a)} \includegraphics[clip,width=0.6\columnwidth]{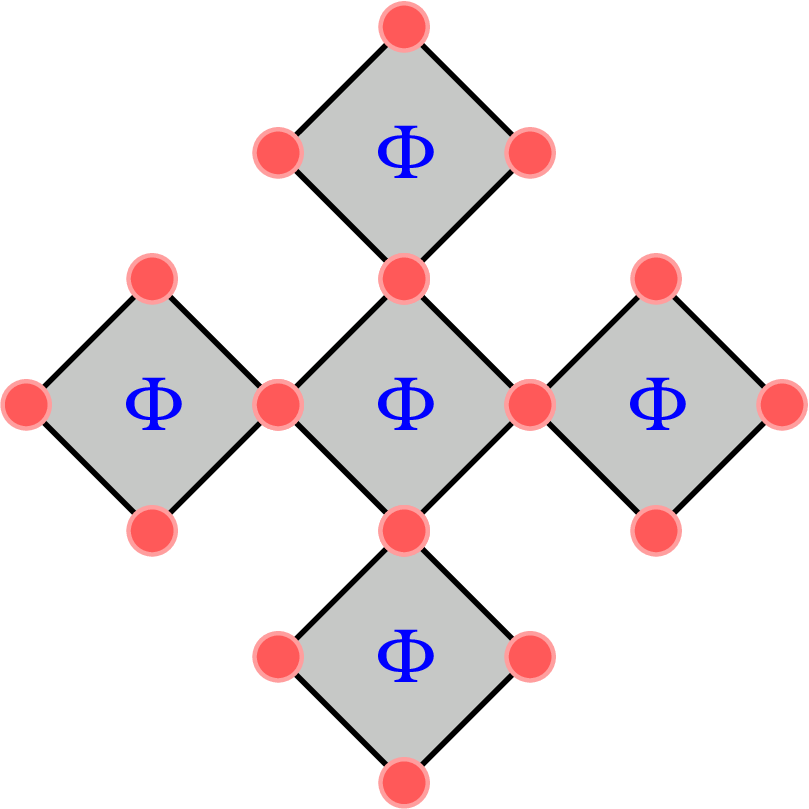}
\textbf{(b)} \includegraphics[clip,width=0.6\columnwidth]{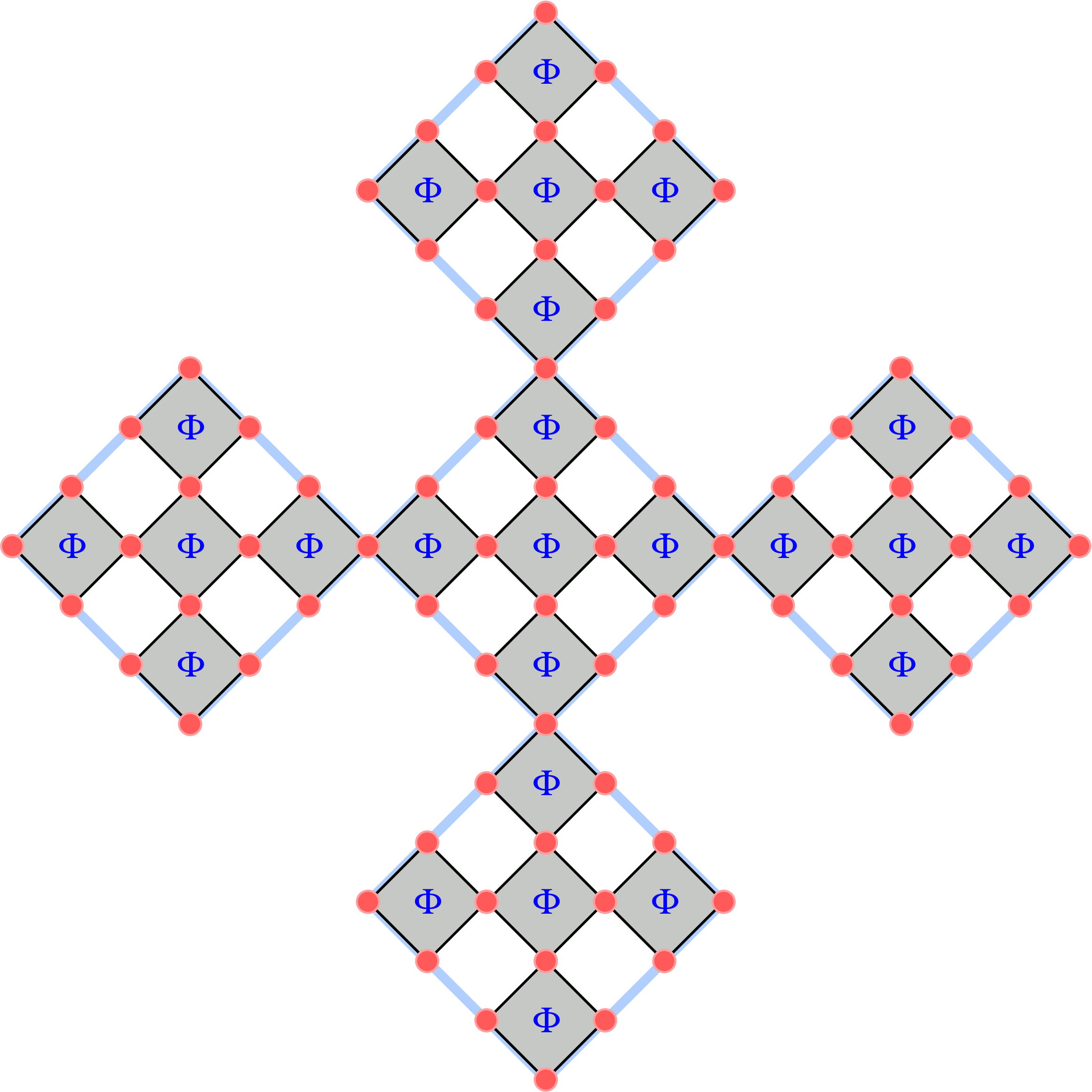}
\textbf{(c)} \includegraphics[clip,width=0.6\columnwidth]{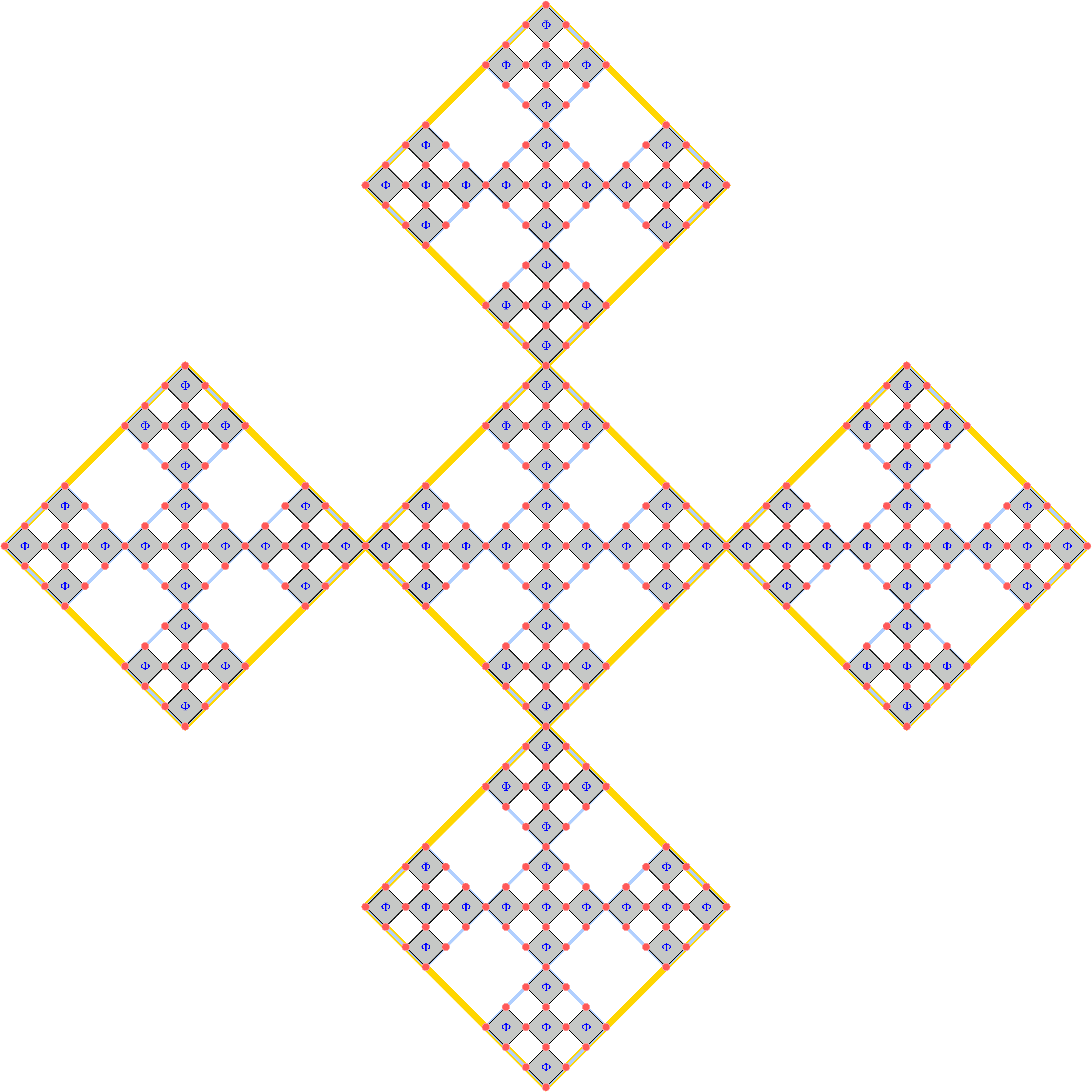}
\caption{Schematic diagram of various generations, \textit{viz.}, 
(a) first generation ($\mathcal{G}_{1}$), 
(b) second generation ($\mathcal{G}_{2}$), and
(c) third generation ($\mathcal{G}_{3}$) of a closed-loop 
Vicsek fractal lattice geometry with all the diamond-shaped closed-loop 
building blocks pierced by a nonzero uniform magnetic flux $\Phi$. }
\label{fig:fractal-growth}
\end{figure*}
%

As of now, the AB caging phenomena has been realized for lattices with 
integer dimensions, such as in 1D and 2D lattices. An intriguing question one 
can ask is that, can we capture this exotic phenomena in a lattice geometry 
with fractional dimension? We address this issue in this research article 
and show that it is indeed possible to realize the AB caging in a lattice 
geometry with \emph{fractional} dimension. In a fractal lattice, as it grows 
in size, we start to have more and more holes appearing in its structure and 
hence it offers \emph{fractional} dimension. This makes fractal lattices very 
interesting and in recent times, it has generated a keen interest in the 
scientific community from various aspects, namely, topological phases of 
matter~\cite{Titus-prb-2018,Shreya-prb-2019,CMsmith-prb-2024,CMsmith-arxiv-2024}, 
quantum Hall effect~\cite{Biplab-prr-2020,Askar-Hall-prb-2020,
CMsmith-Hall-prr-2020}, flatband physics~\cite{Pal-Saha-prb-2018,
Atanu-phys-scr-2021,Hahafi-fractal-FB, Chen-fractal-FB-1,
Chen-fractal-FB-2,Sougata-fractal-FB}, Bose-Einstein 
condensates~\cite{Anna-pra-2024}, Josephson effect~\cite{Vladimir-apl-2024} 
etc., are to name a few among others. One of 
the recent exciting developments is that, these complex fractal structures are 
not just anymore mere theoretical artifact, but they have been synthesized in 
real-life experiments in the laboratory using various advanced technologies, 
such as self-assembly molecular growth techniques~\cite{self-assembly-1,
self-assembly-2}, atomic manipulation method with scanning tunnelling 
microscope~\cite{CMsmith-nat-phys-2019,CMsmith-nat-phys-2024}, and femtosecond 
laser-writing technique~\cite{Hahafi-fractal-FB,Chen-fractal-FB-1,
Chen-fractal-FB-2}. These experimental advances in the fractal growth 
techniques encourage us to theoretically investigate the AB caging phenomena 
in a quantum fractal geometry~\cite{quantum-fractal-nat-phys}. 

In this study, we have considered a fractal lattice geometry built out of 
diamond-shaped building blocks which is popularly known as the 
\emph{Vicsek} snowflake or \emph{Vicsek} 
fractal~\cite{spacecraft-Vicsek-fractal,harmonic-fn-Vicsek-fractal,
random-walk-Vicsek-fractal,pal-prb-2012,pal-pra-2013} with the 
Hausdorff dimension $d_{f} = \log(5)/\log(3) \approx 1.465$. 
Such fractal geometry is not only important from the point view 
of physics, but also it is found to be very useful in the field of 
aerospace engineering~\cite{spacecraft-Vicsek-fractal}, 
mathematics~\cite{harmonic-fn-Vicsek-fractal}, 
random walk statistics~\cite{random-walk-Vicsek-fractal}, and so on. 
In our model, all the diamond-shaped closed-loop plaquettes are penetrated 
by a nonzero uniform magnetic flux as depicted in 
Fig.~\ref{fig:fractal-growth}. Due to the quantum mechanical effect, 
this will induce a geometric phase factor known as the AB phase in the 
wavefunction of the electron, when it will traverse through this lattice 
structure. We have shown that, for a special value of the magnetic 
flux, one can completely block the transmission of the electron in this 
fractal system. Thus, we encounter the phenomenon of AB caging in such a 
quantum fractal geometry. We quantify this result by a rigorous computation 
of the energy spectrum, density of states, two-terminal transport 
characteristics, and persistent current upto the third generation Vicsek 
fractal geometry. We have also verified that the phenomenon of AB caging in 
this fractal geometry is very robust against the onsite disorder in the system. 

Our findings and analysis are presented in the following. The structure 
of the article is as follows: In Sec.~\ref{sec:energy-spectrum}, we define 
the tight-binding Hamiltonian which describe our lattice model, and show the 
results for the energy eigenvalue spectrum as a function of the magnetic flux 
for this fractal lattice geometry in its various generations. 
Sec.~\ref{sec:DOS-and-transport} deals with the results and analysis for 
the density of states and the two-terminal transport properties of the system 
with different values of the magnetic flux upto the third generation Vicsek 
fractal geometry, alongwith a brief discussion on the effect of onsite 
disorder on the AB caging phenomenon for this fractal system. 
The results for the persistent current in this closed-loop 
fractal geometry is presented in the Sec.~\ref{sec:persistent-current}. Finally, 
in Sec.~\ref{sec:conclusion}, we draw our conclusion by summarizing the central 
result of this study and possible future scope.  
\section{The Hamiltonian and the energy spectrum}
\label{sec:energy-spectrum} 
The single-particle states in a Vicsek fractal lattice as shown in 
Fig.~\ref{fig:fractal-growth} can be mathematically described by the following 
tight-binding Hamiltonian:
\begin{equation}
\bm{H} = \sum_{n} \varepsilon_{n}c_{n}^{\dagger}c_{n} + 
\sum_{\langle n,m\rangle} \big(t_{nm}e^{i\theta_{nm}}c_{m}^{\dagger}c_{n} 
+ \textrm{h.c.}\big),
\label{eq:hamiltonian}
\end{equation}
where the first term represents the potential energy term and the second term 
represents the kinetic energy term. In the language of tight-binding formalism, 
$\varepsilon_{n}$ is known as the onsite potential at the $n$th site 
and $t_{nm}$ is known as the nearest-neighbor hopping integral between 
the two neighboring sites `$n$' and `$m$' in the fractal lattice geometry. 
$c_{n}^{\dagger}\,(c_{n})$ are the operators which creates (annihilates) a 
single-particle state at the $n$th site. 
$\theta_{nm} = \frac{2\pi\Phi}{4\Phi_{0}}$ is the AB phase acquired by 
the electron wavefunction while hopping from the $n$th site to $m$th site 
along a diamond-shaped loop in the clockwise direction as a effect 
of the enclosed magnetic flux $\Phi$ (measured in units of the fundamental 
flux quantum $\Phi_{0} = h/e$). The dimension of the Hamiltonian for the 
system is governed by the total number of sites in the Vicsek fractal lattice 
geometry in a given generation $\mathcal{G}_{\ell}$ following the 
prescription given by  
\begin{equation}
\mathcal{N}_{\ell} = \big[3R^{\ell} + 1 \big],
\label{eq:growth-rule} 
\end{equation}
where $R=5$ is the repetition factor for the Vicsek fractal geometry and 
`$\ell$' represents the generation index.  
%
\begin{figure*}[ht]
\textbf{(a)} \includegraphics[clip,width=0.6\columnwidth]{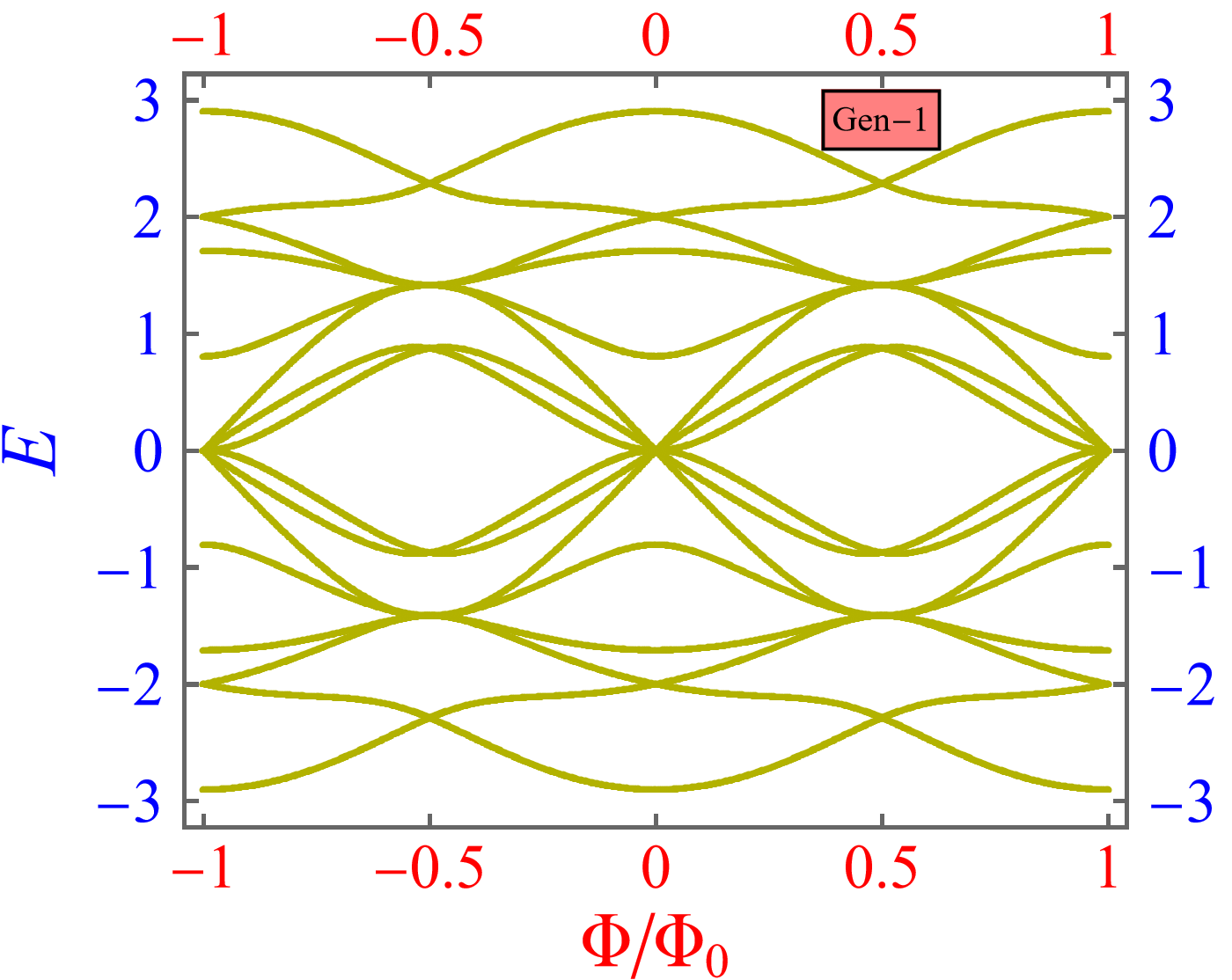}
\textbf{(b)} \includegraphics[clip,width=0.6\columnwidth]{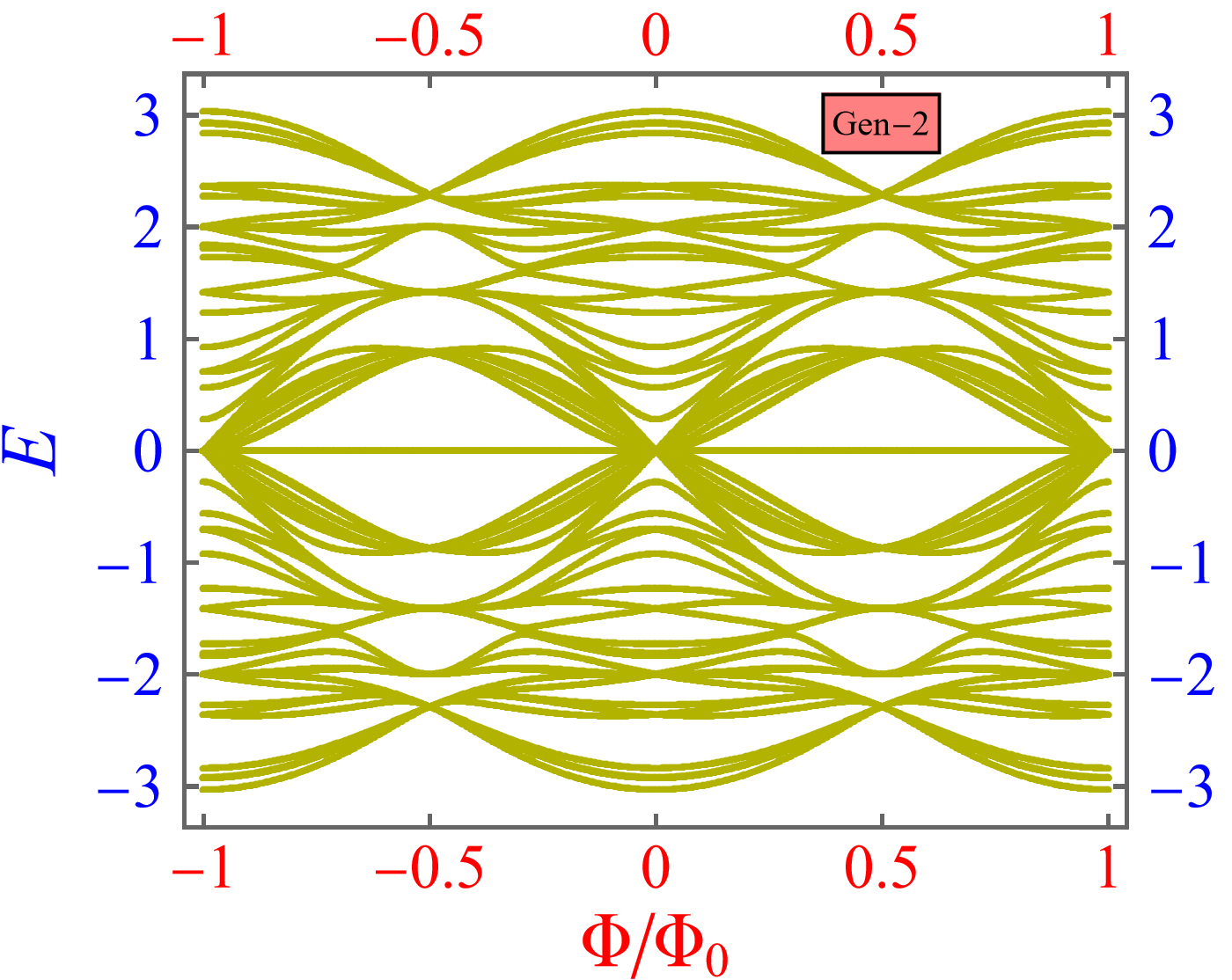}
\textbf{(c)} \includegraphics[clip,width=0.6\columnwidth]{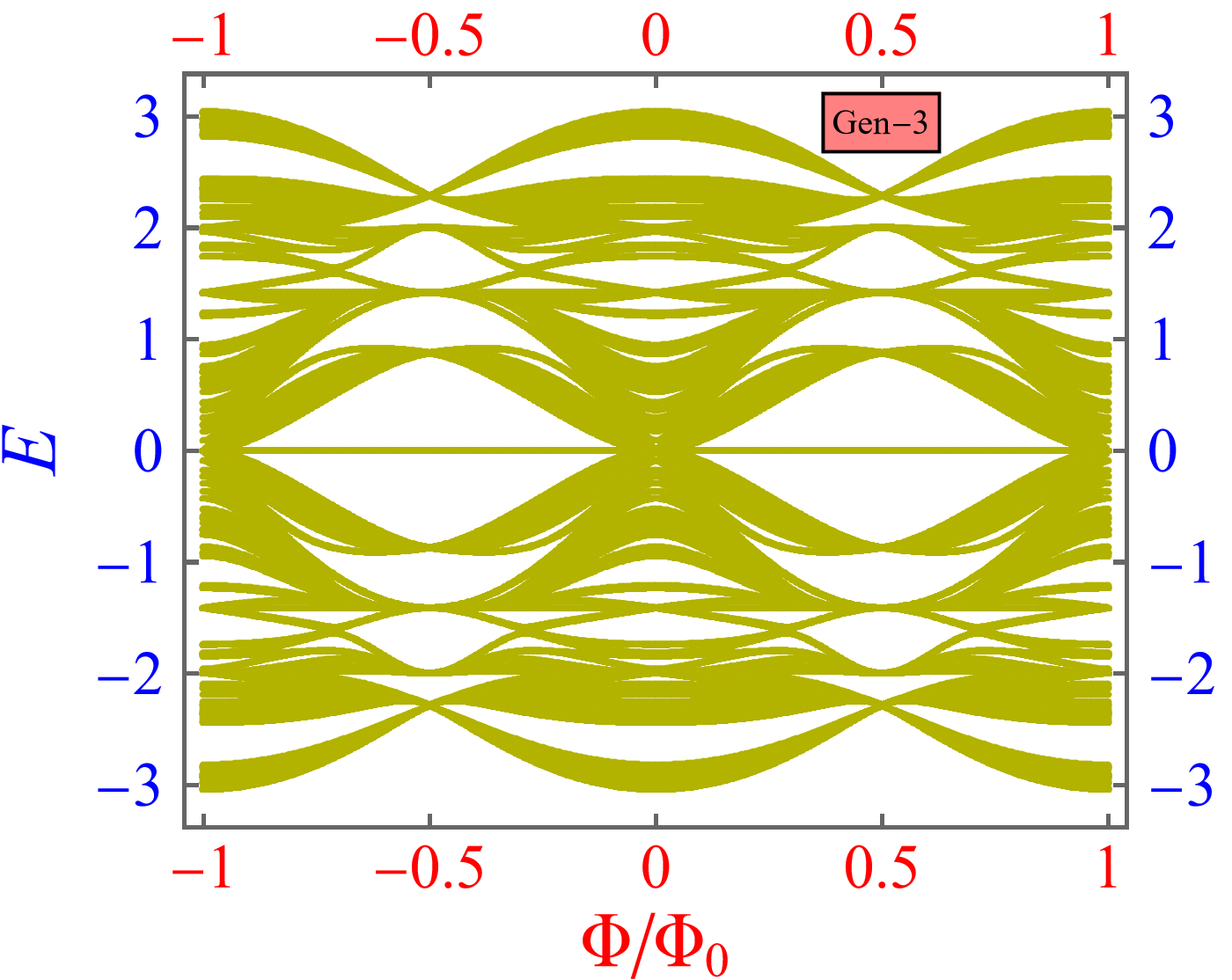}
\caption{Variation in the energy eigenvalue spectrum ($E$) 
as function of the Aharonov-Bohm flux ($\Phi$) for 
(a) first generation (with $\mathcal{N}_{1}=16$ sites), 
(b) second generation (with $\mathcal{N}_{2}=76$ sites), and
(c) third generation (with $\mathcal{N}_{3}=376$ sites) 
Vicsek fractal lattice structure, respectively. We set the onsite 
potential $\varepsilon_{n}=0$ for all $n$-sites and the nearest-neighbor 
hopping integral $t_{nm}=t=1$ for all $\langle n,m\rangle$ bonds.}
\label{fig:eigenvalue-spectrum}
\end{figure*}
%

All the important information and properties of the Vicsek fractal system 
can be extracted from the Hamiltonian of the system. At first, we focus on 
the energy eigenvalues spectrum of the Vicsek fractal system. One can easily 
obtain the energy eigenvalue spectrum for any $\ell$th generation Vicsek 
fractal geometry by diagonalizing the corresponding 
$(\mathcal{N}_{\ell} \times \mathcal{N}_{\ell})$ Hamiltonian matrix for the 
system. We have done this upto the third generation Vicsek fractal lattice 
and shown the variation in the energy eigenvalue spectrum for the system as 
a function of the uniform Aharonov-Bohm flux $\Phi$ 
(see Fig.~\ref{fig:eigenvalue-spectrum}). The Aharonov-Bohm flux $\Phi$ has 
a periodicity $0$ to $\Phi_{0}$ (the fundamental flux quantum), which is 
clearly reflected from the energy eigenvalue pattern displayed in 
Fig.~\ref{fig:eigenvalue-spectrum}. The energy eigenvalues $E$ are measured 
in units of the hopping integral $t$. One of the fundamental features of 
any fractal system is that, it has an inherent self-similar character in it, 
which is also emulated in the energy spectrum shown in 
Fig.~\ref{fig:eigenvalue-spectrum}(a)-(c). As the fractal generation increases, 
the number of lattice sites in the system also increases. 
Hence, the energy spectrum also become more and more denser, and after a 
certain generation it reaches the saturation value where one cannot distinguish 
the spectrum from its previous generation. 

If we scan through the $x$-axis in the Fig.~\ref{fig:eigenvalue-spectrum}(a)-(c), 
we can identify a very interesting thing happening at the value 
$\Phi=\pm \Phi_{0}/2$ --- at this particular value of the magnetic flux, 
the energy spectrum shrink to a few points. This is a clear indication of the 
fact that, at this particular value of the magnetic flux, the single-particle 
states in this fractal lattice model are highly localized giving rise to the 
AB caging phenomena in a quantum fractal system. Another interesting 
observation is that, starting from the second generation fractal lattice 
onward, the eigenvalue at $E=0$ emerges in the spectrum (see 
Fig.~\ref{fig:eigenvalue-spectrum}(b)), which is not there in the spectrum 
for the first generation fractal system (see 
Fig.~\ref{fig:eigenvalue-spectrum}(a)). 
The possible reason behind the appearance of this feature is that the 
eigenvalue at $E=0$ acts like an edge or corner state for this particular 
fractal lattice geometry. As we go the second generation Vicsek fractal 
system (see Fig.~\ref{fig:fractal-growth}(b)), the holes in the system 
becomes more prominent, clearly distinguishing the notion of 
the edge or corner in the fractal system. 
It is to be noted that, the $E=0$ state is very robust 
against the variation of the magnetic flux in the system. 
In Sec.~\ref{sec:DOS-and-transport}, we substantiate the results and analysis 
presented in this section through a rigorous computation of the density of 
states and the two-terminal transport properties for this fractal lattice 
model.                
\section{The density of states and the transport characteristics}
\label{sec:DOS-and-transport} 
The density of states (DOS) of a tight-binding lattice model gives us an 
indication about the character of the single-particle eigenfunctions 
in the system, \textit{i.e.}, whether they are extended or localized. 
We use the Green's function formalism to evaluate the DOS for such 
closed-loop fractal structures~\cite{ac-prb-2005,pal-epjb-2012,nandy-jpcm-2015}. 
The DOS in terms of the Green's function $\bm{G}(E,\Phi)$ of the system 
is given by 
\begin{equation}
\mathfrak{D}(E,\Phi) = -\dfrac{1}{\pi\mathcal{N_{\ell}}}\textrm{Im} 
\Big[ \textrm{Tr}\left[\bm{G}(E,\Phi)\right]\Big],
\label{eq:DOS}
\end{equation} 
where $\bm{G}(E,\Phi) = \big[(E + i\eta){\bm I} - 
\bm{H} \big]^{-1}$ (with $\eta \rightarrow 0^{+}$), ${\bm I}$ is an 
Identity matrix of dimension $(\mathcal{N_{\ell}} \times \mathcal{N_{\ell}})$, 
$\mathcal{N_{\ell}}$ is the number of sites in an 
$\ell$th generation Vicsek fractal lattice, and `$\textrm{Tr}$' represents the 
trace of the Green's function matrix.   
%
\begin{figure*}[ht]
\textbf{(a)} \includegraphics[clip,width=0.6\columnwidth]{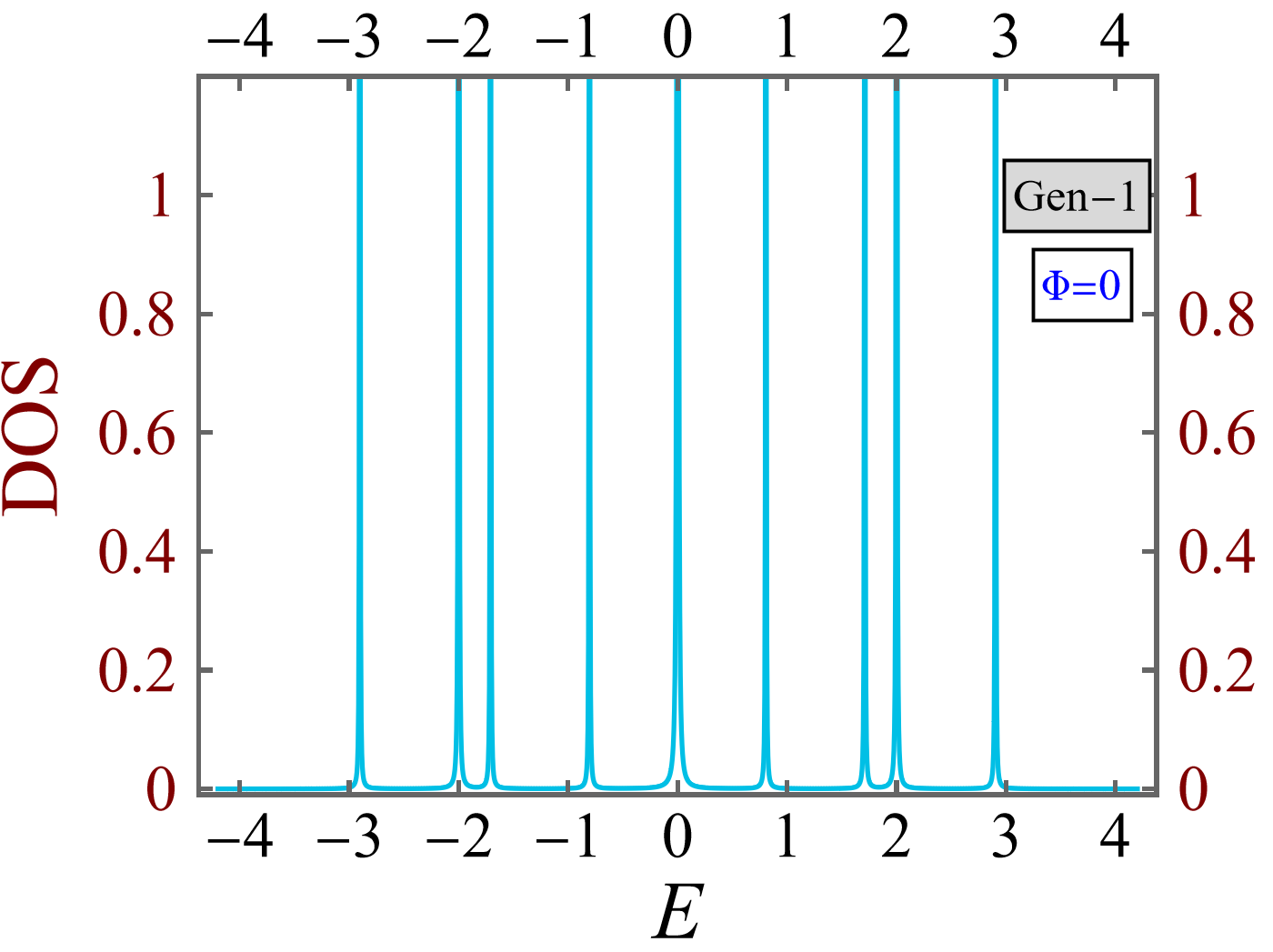}
\textbf{(b)} \includegraphics[clip,width=0.6\columnwidth]{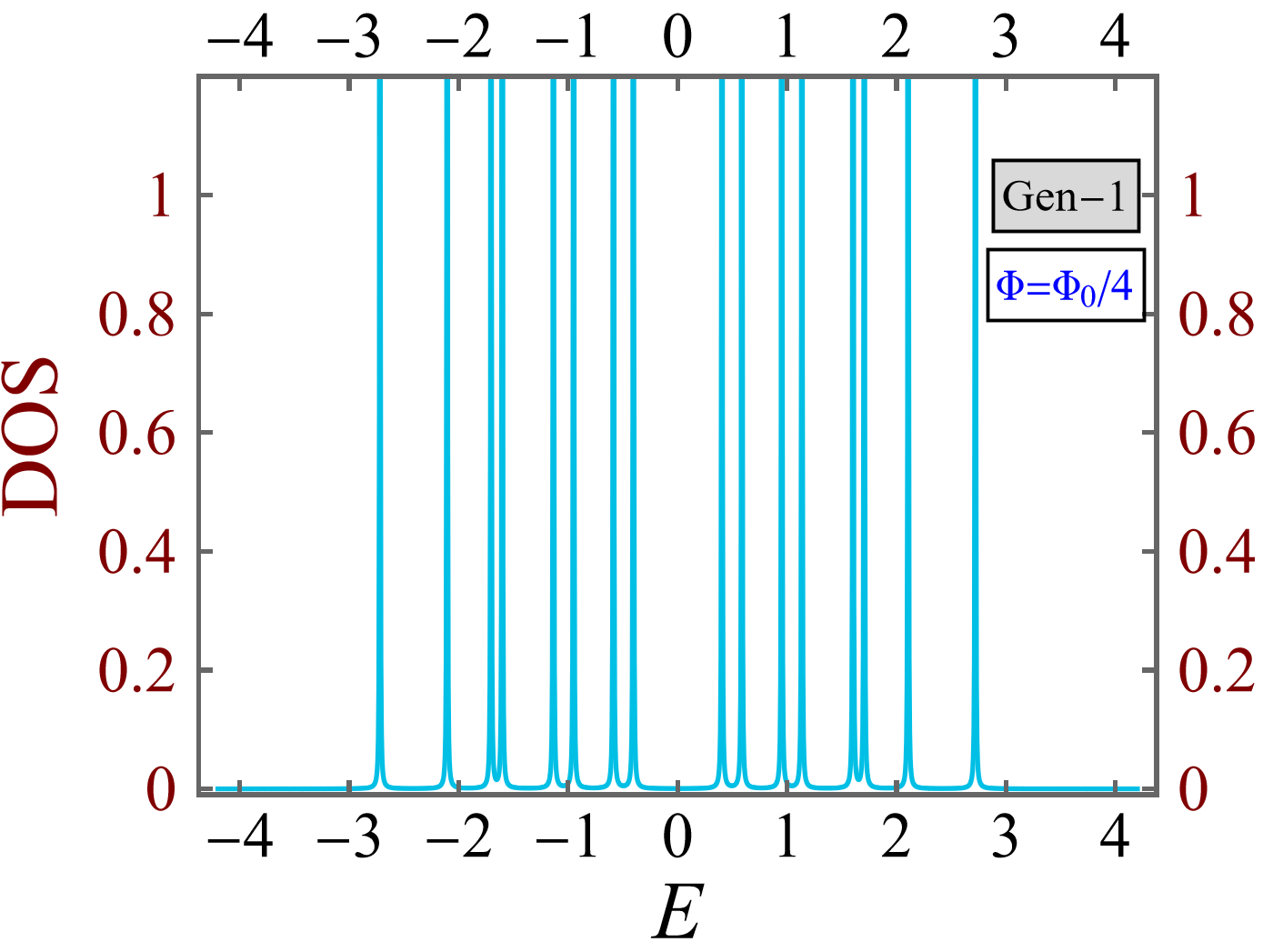}
\textbf{(c)} \includegraphics[clip,width=0.6\columnwidth]{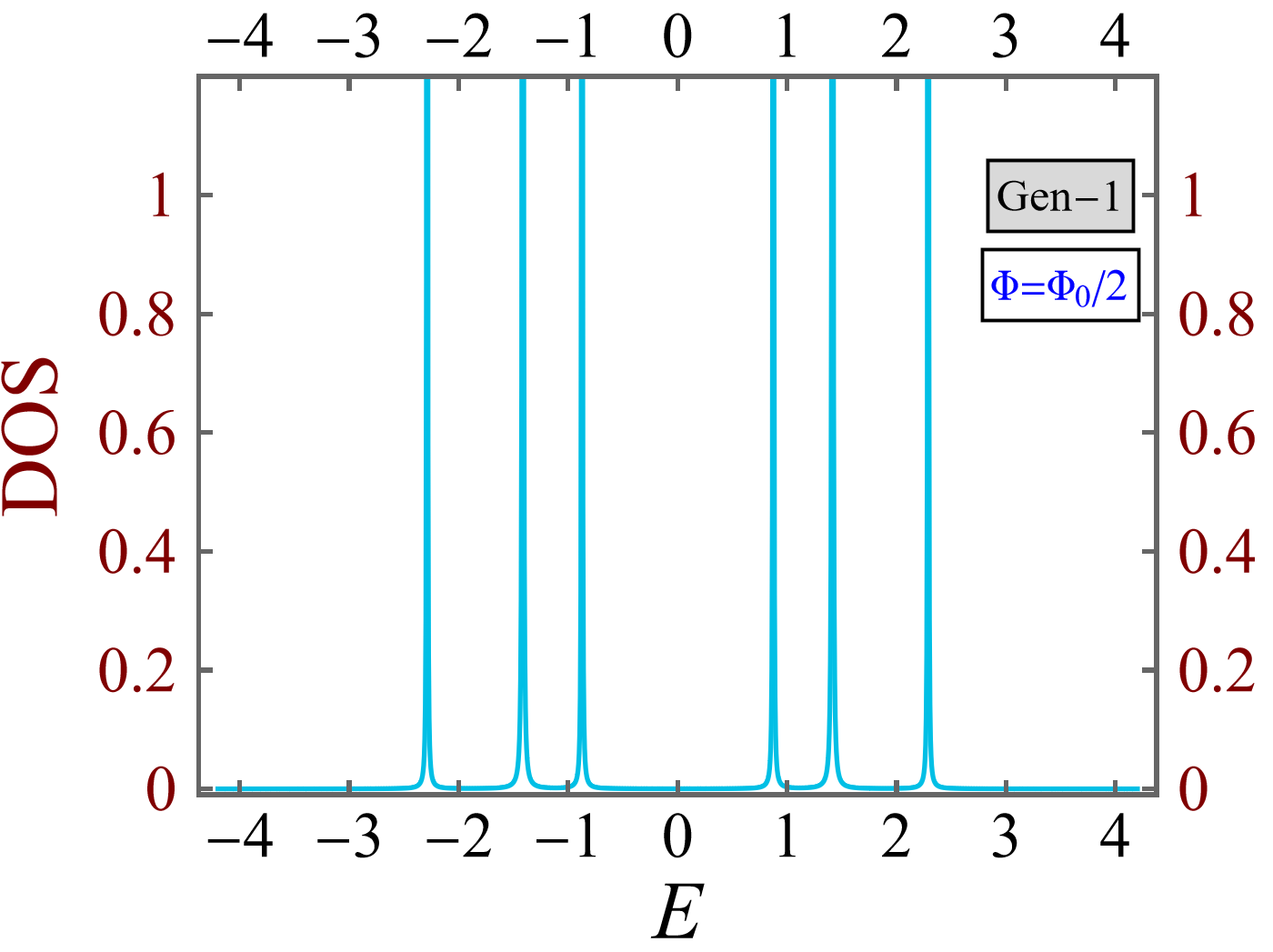}
\vskip 0.1cm
\textbf{(d)} \includegraphics[clip,width=0.6\columnwidth]{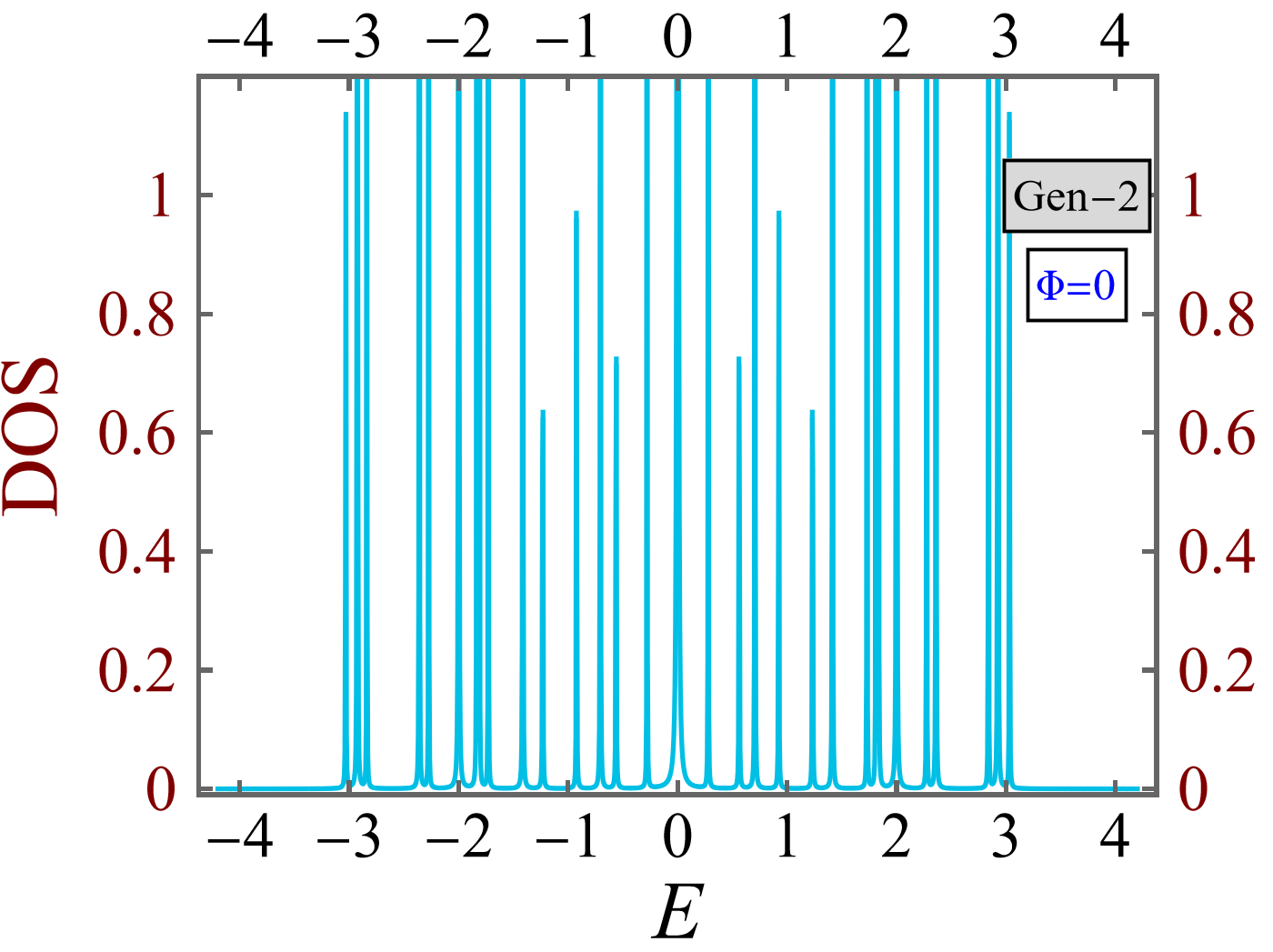}
\textbf{(e)} \includegraphics[clip,width=0.6\columnwidth]{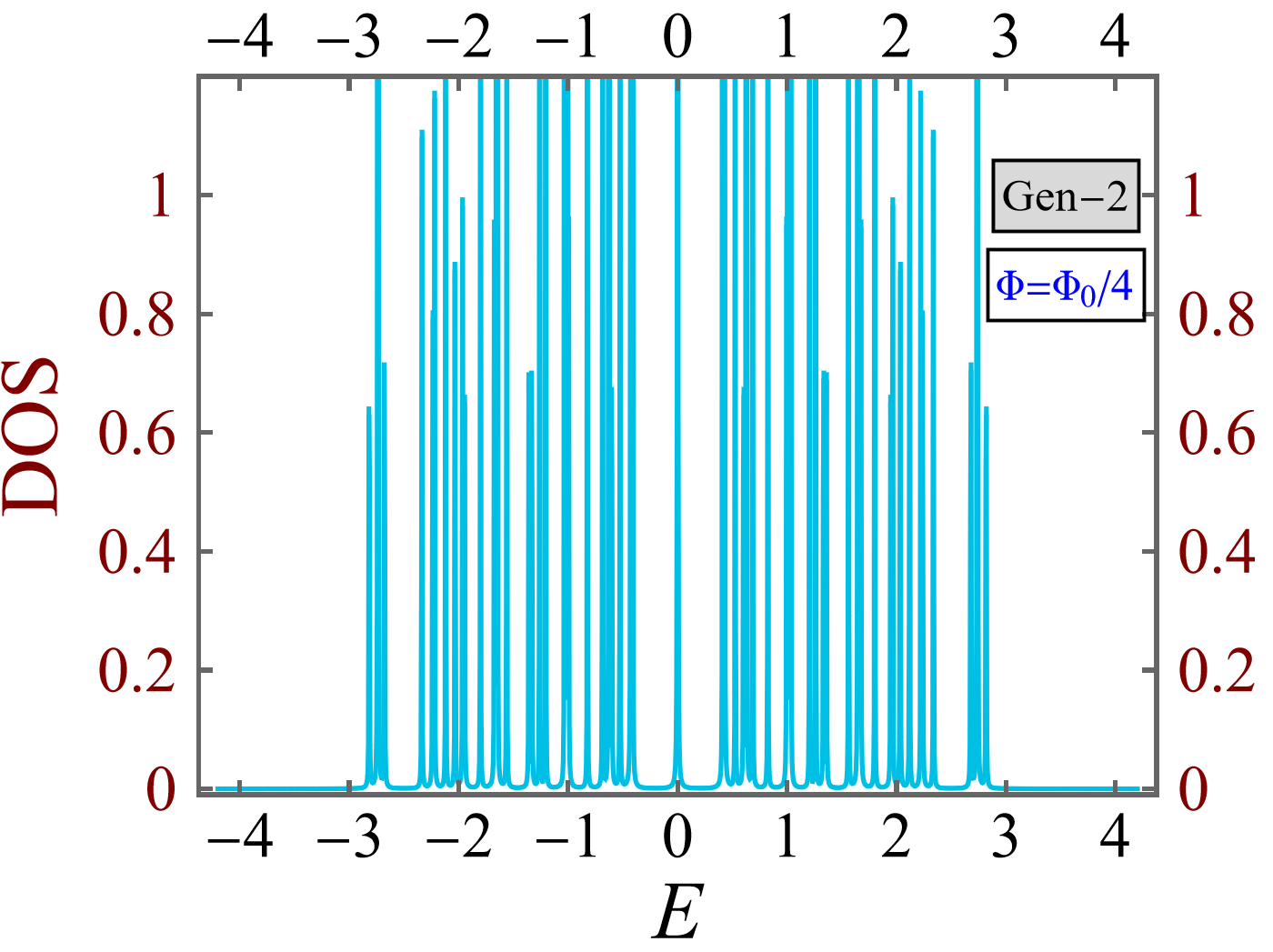}
\textbf{(f)} \includegraphics[clip,width=0.6\columnwidth]{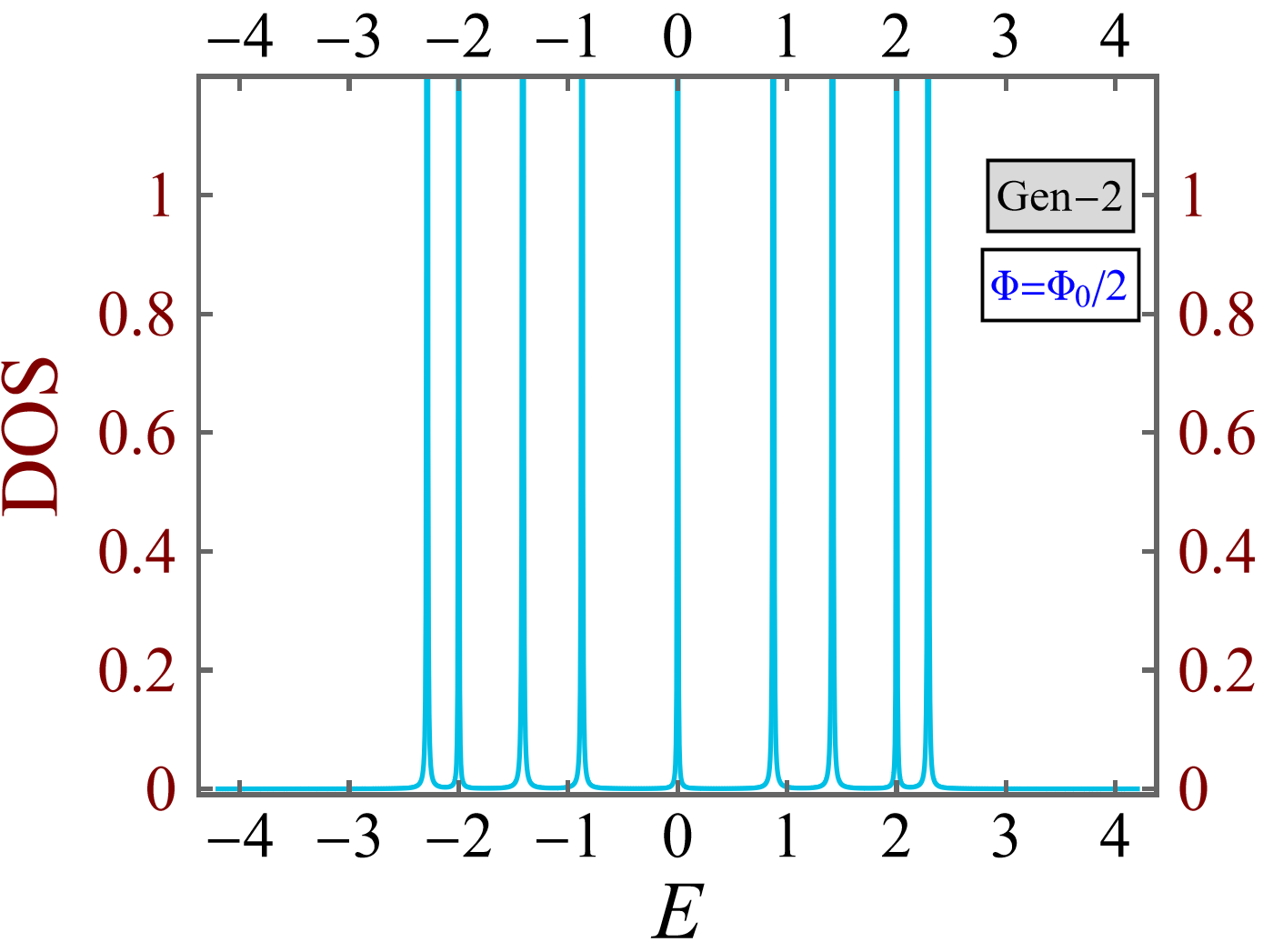}
\vskip 0.1cm
\textbf{(g)} \includegraphics[clip,width=0.6\columnwidth]{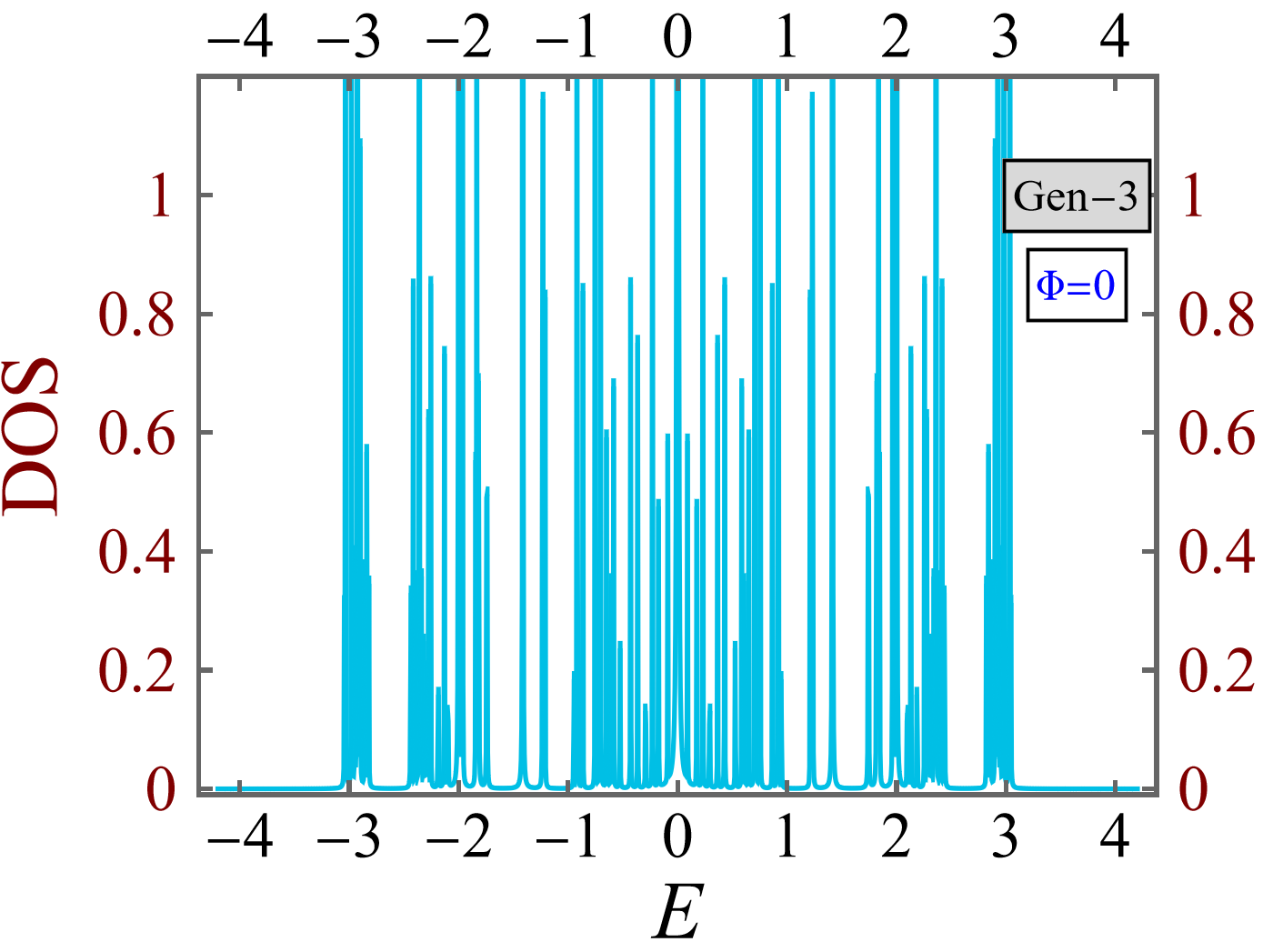}
\textbf{(h)} \includegraphics[clip,width=0.6\columnwidth]{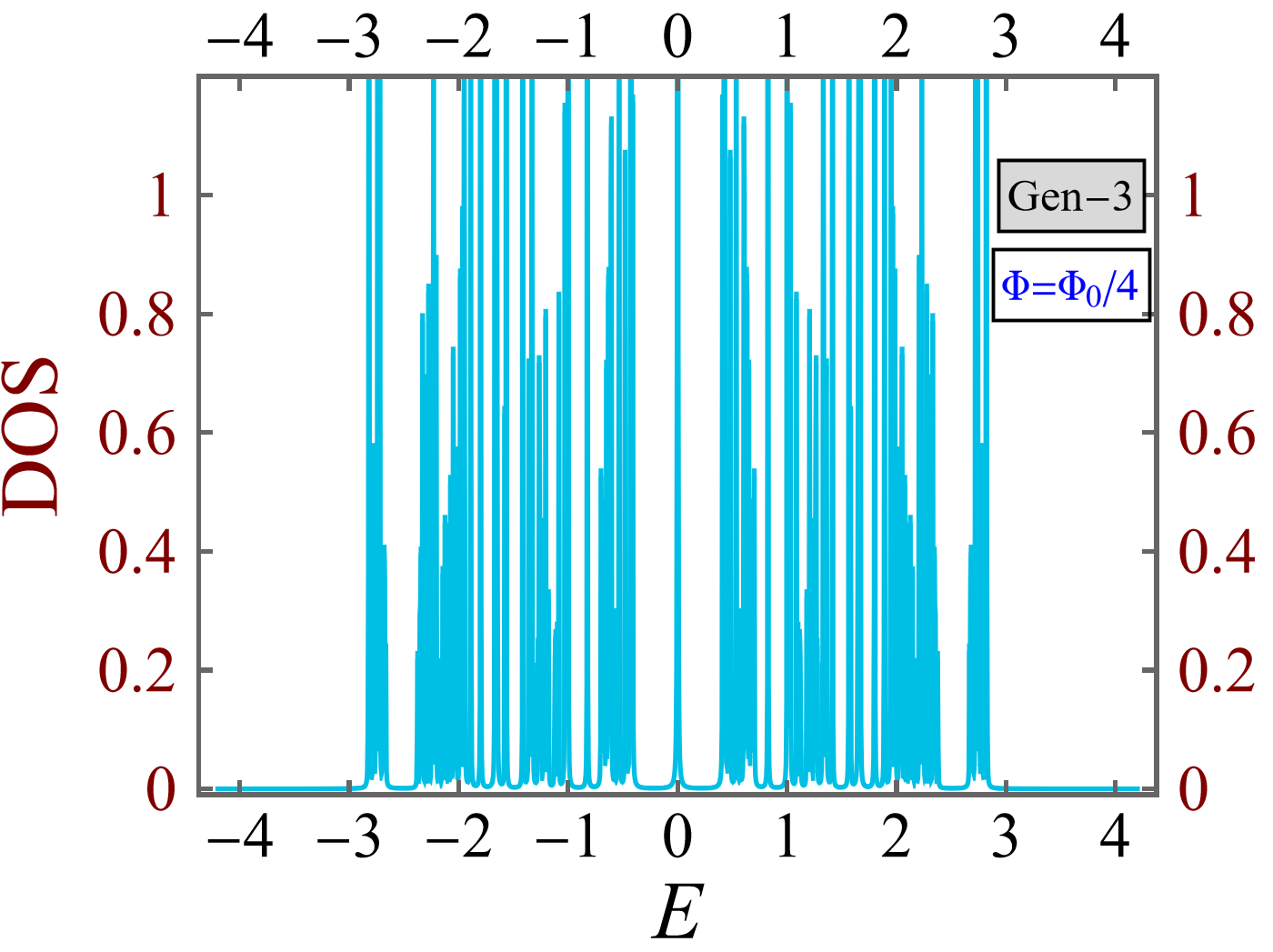}
\textbf{(i)} \includegraphics[clip,width=0.6\columnwidth]{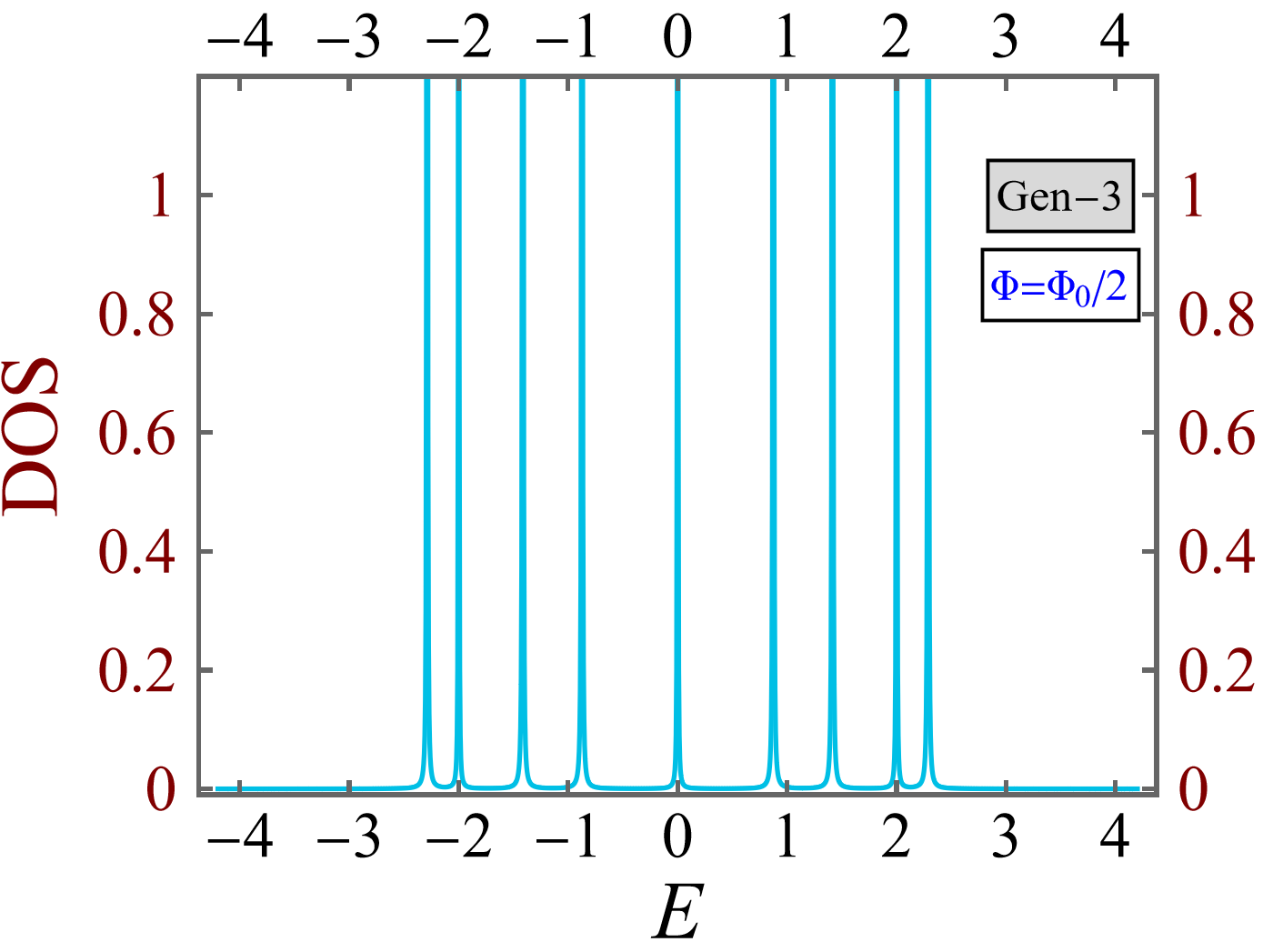}
\caption{The plots for the density of states (DOS) 
as a function of the energy ($E$) of the electron for the 
(a)-(c) first generation ($\mathcal{G}_{1}$), 
(d)-(f) second generation ($\mathcal{G}_{2}$), and 
(g)-(i) third generation ($\mathcal{G}_{3}$) Vicsek fractal lattice. 
The left column is for $\Phi=0$, 
the middle column is for $\Phi=\Phi_{0}/4$, and 
the right column is for $\Phi=\Phi_{0}/2$.}
\label{fig:DOS}
\end{figure*}
%

Using the above formalism, we have computed the DOS for the Vicsek fractal 
lattice in its different generations as depicted in Fig.~\ref{fig:DOS}. In 
general, for fractal lattices, the DOS display a multifractal character 
with the single-particle states are being localized following a power-law 
localization pattern~\cite{ac-prb-2005,pal-epjb-2012,nandy-jpcm-2015}. 
This is what we observe in the DOS spectrum for the Vicsek fractal lattice 
model as shown in Fig.~\ref{fig:DOS}. As we are interested in the effect 
of the magnetic flux on the single-particle states in this fractal lattice, therefore we have illustrated the DOS for three 
different values of the magnetic flux, 
\textit{viz.}, $\Phi=0$, $\Phi_{0}/4$ and $\Phi_{0}/2$, respectively, for the 
first generation ($\mathcal{G}_{1}$) [see Fig.~\ref{fig:DOS}(a)-(c)], 
second generation ($\mathcal{G}_{2}$) [see Fig.~\ref{fig:DOS}(d)-(f)],
and third generation ($\mathcal{G}_{3}$) [see Fig.~\ref{fig:DOS}(g)-(i)] 
Vicsek fractal lattice. We note that, as the system size grows, we have 
more denser packing in the DOS spectrum as more number of states are available 
for the electron to occupy. However, if we focus on the extreme right column in 
Fig.~\ref{fig:DOS}, we observe that for $\Phi=\Phi_{0}/2$, the DOS spectrum 
for all the three generations $\mathcal{G}_{1}$, $\mathcal{G}_{2}$, and 
$\mathcal{G}_{3}$, respectively, contracted to only a few sharp 
lines indicating towards the highly localized 
states. This clearly corroborates with the results for the energy spectrum 
depicted in the previous section and confirms the phenomenon of AB caging 
for this fractal lattice geometry.    
%
\begin{figure*}[ht]
\textbf{(a)} \includegraphics[clip,width=0.6\columnwidth]{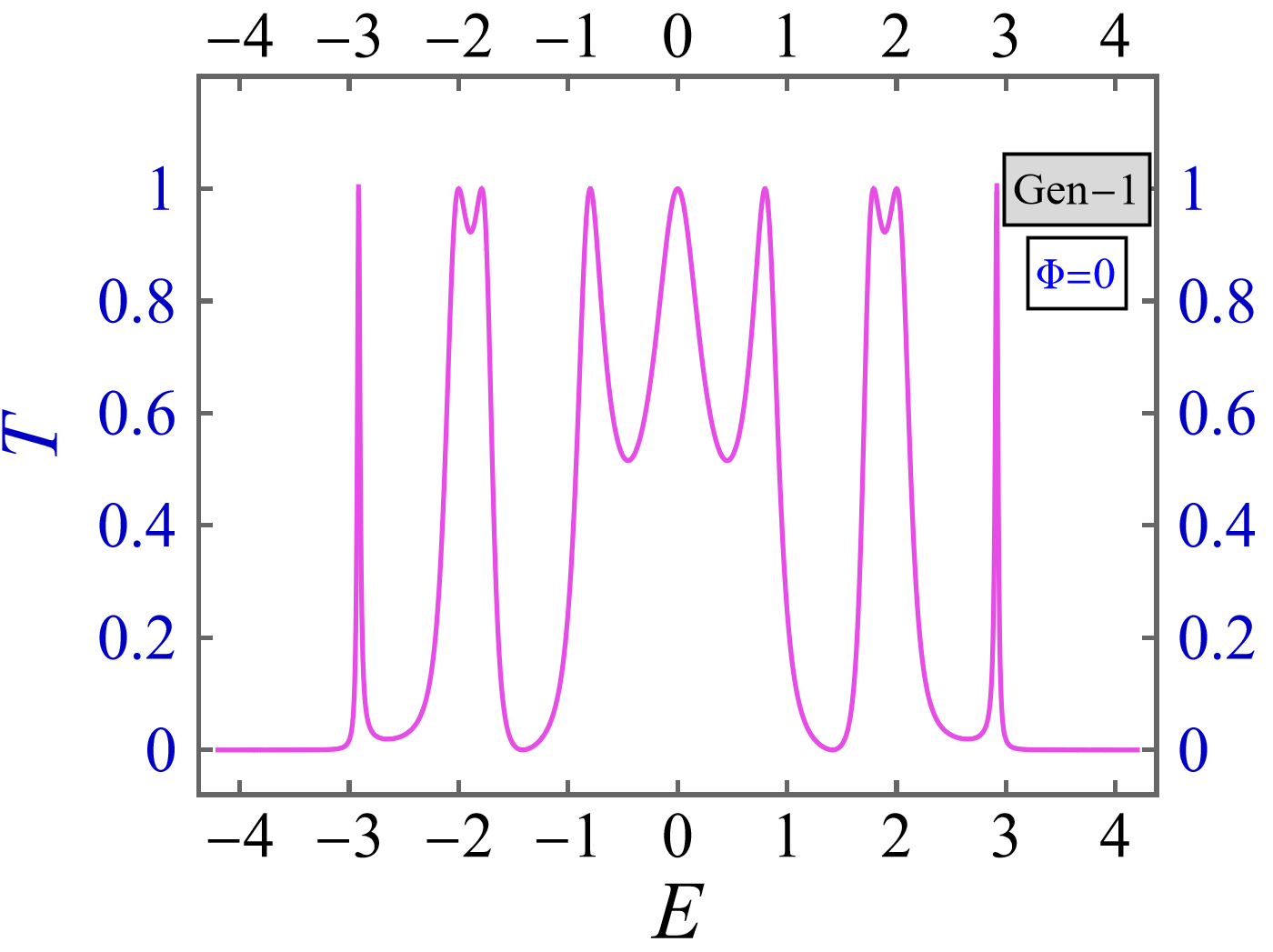}
\textbf{(b)} \includegraphics[clip,width=0.6\columnwidth]{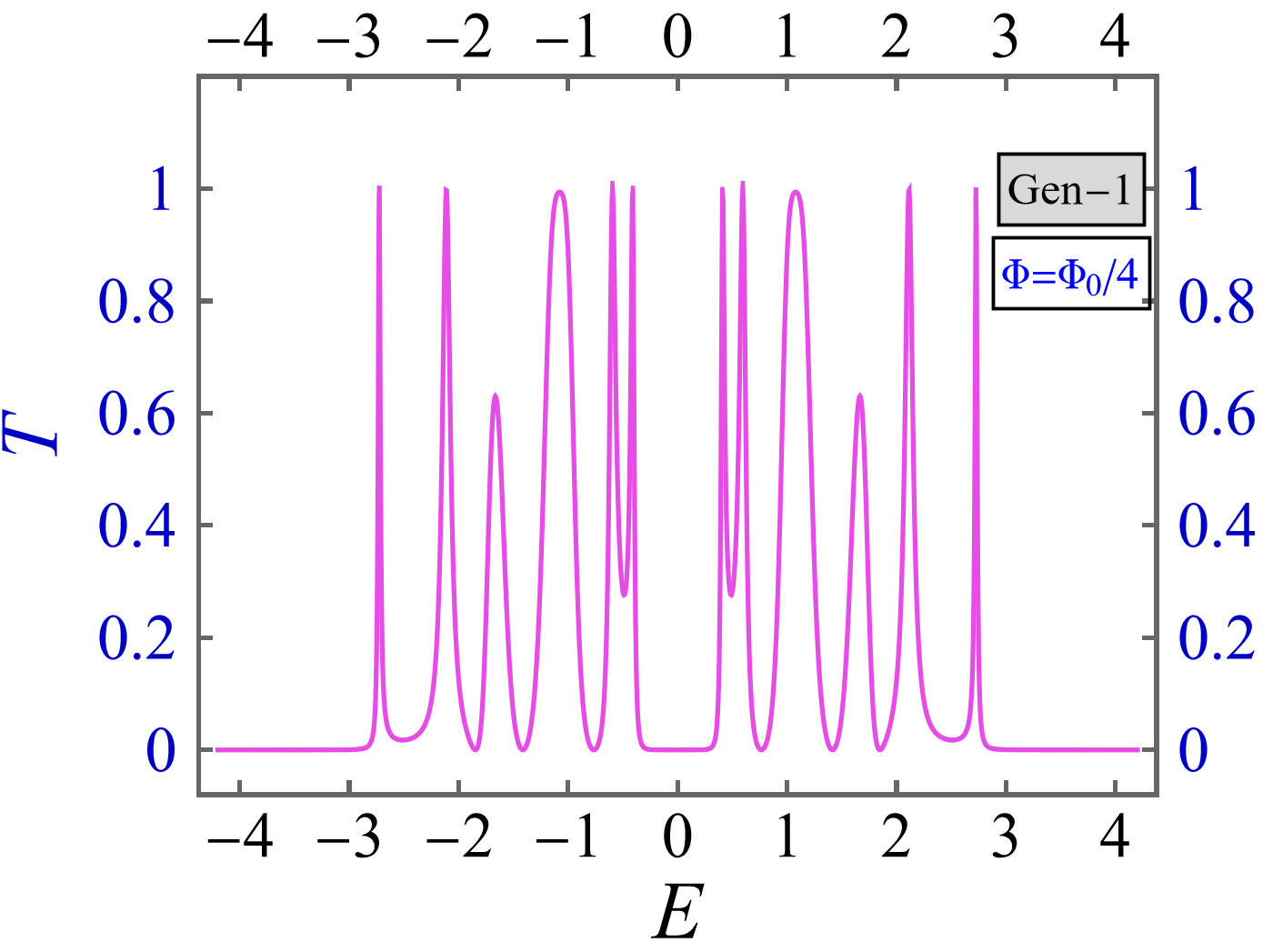}
\textbf{(c)} \includegraphics[clip,width=0.6\columnwidth]{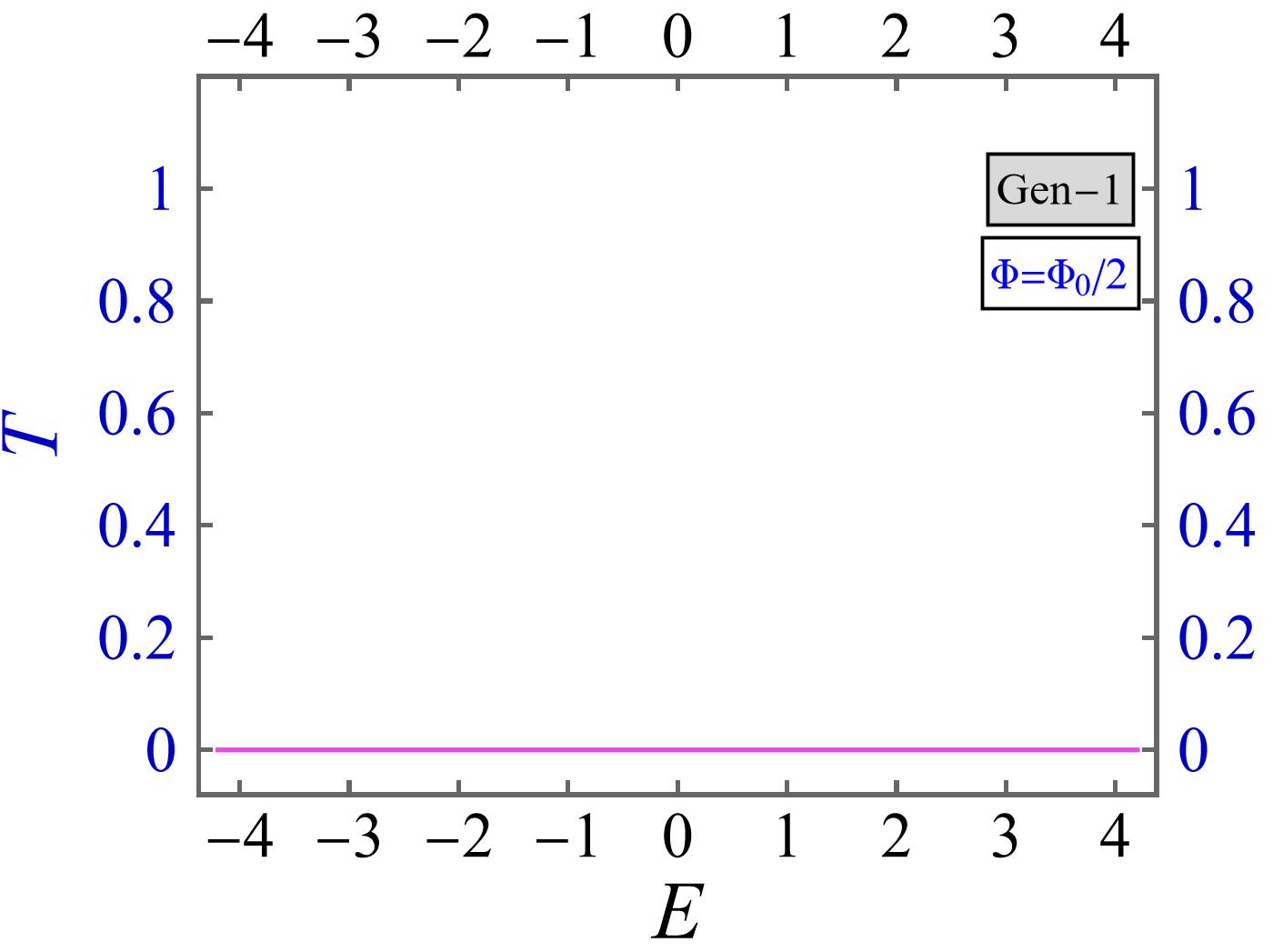}
\vskip 0.1cm
\textbf{(d)} \includegraphics[clip,width=0.6\columnwidth]{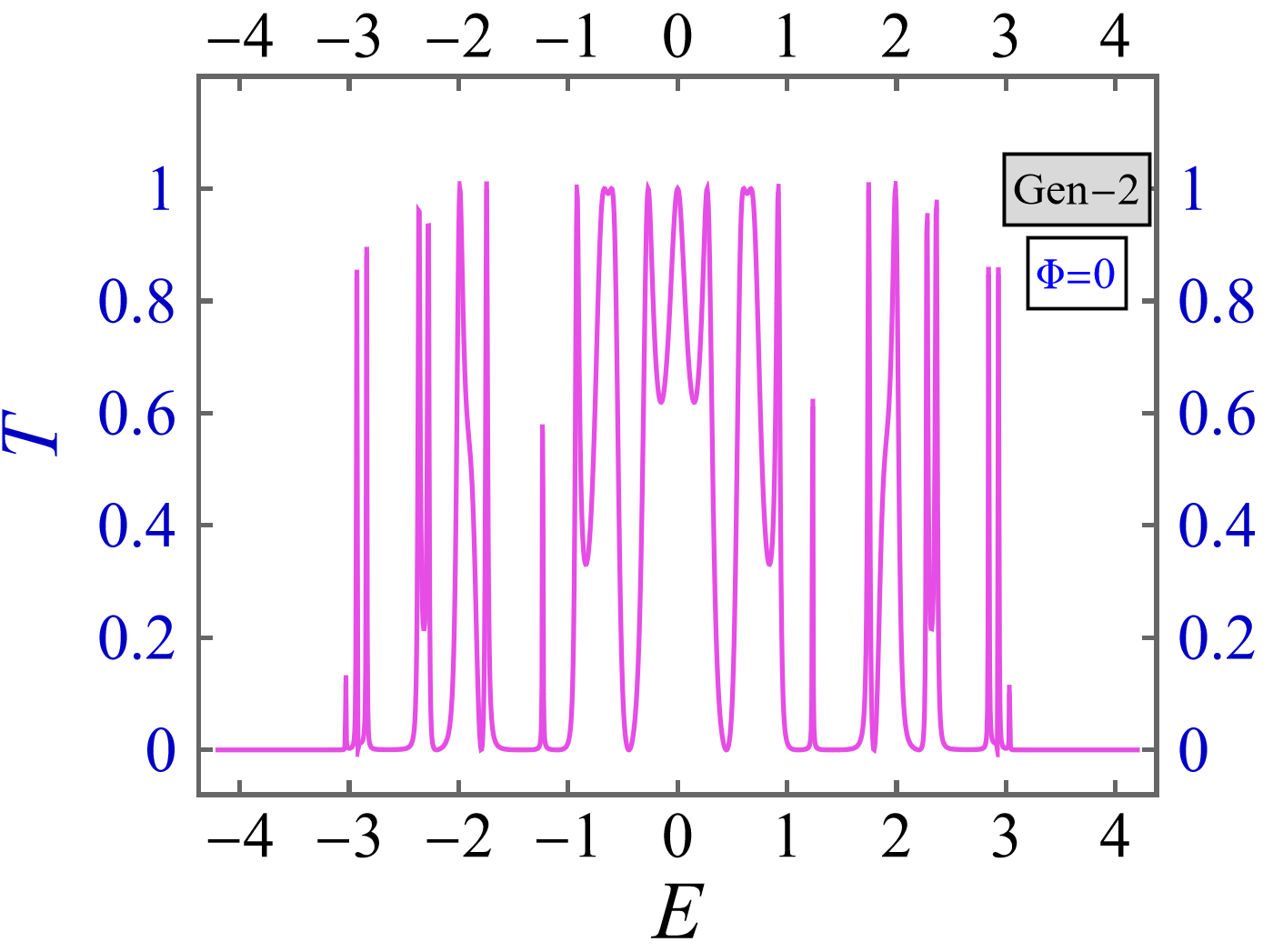}
\textbf{(e)} \includegraphics[clip,width=0.6\columnwidth]{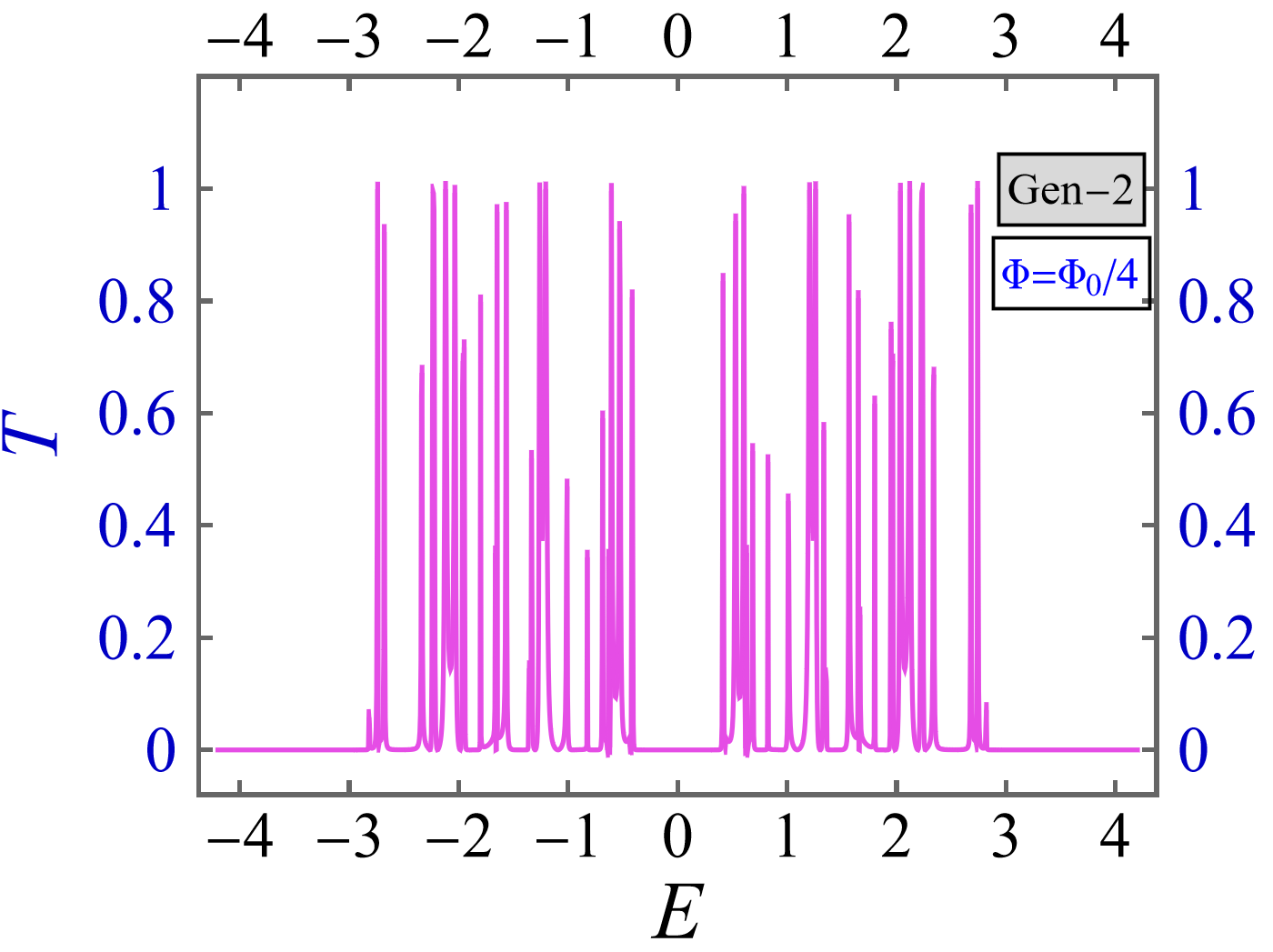}
\textbf{(f)} \includegraphics[clip,width=0.6\columnwidth]{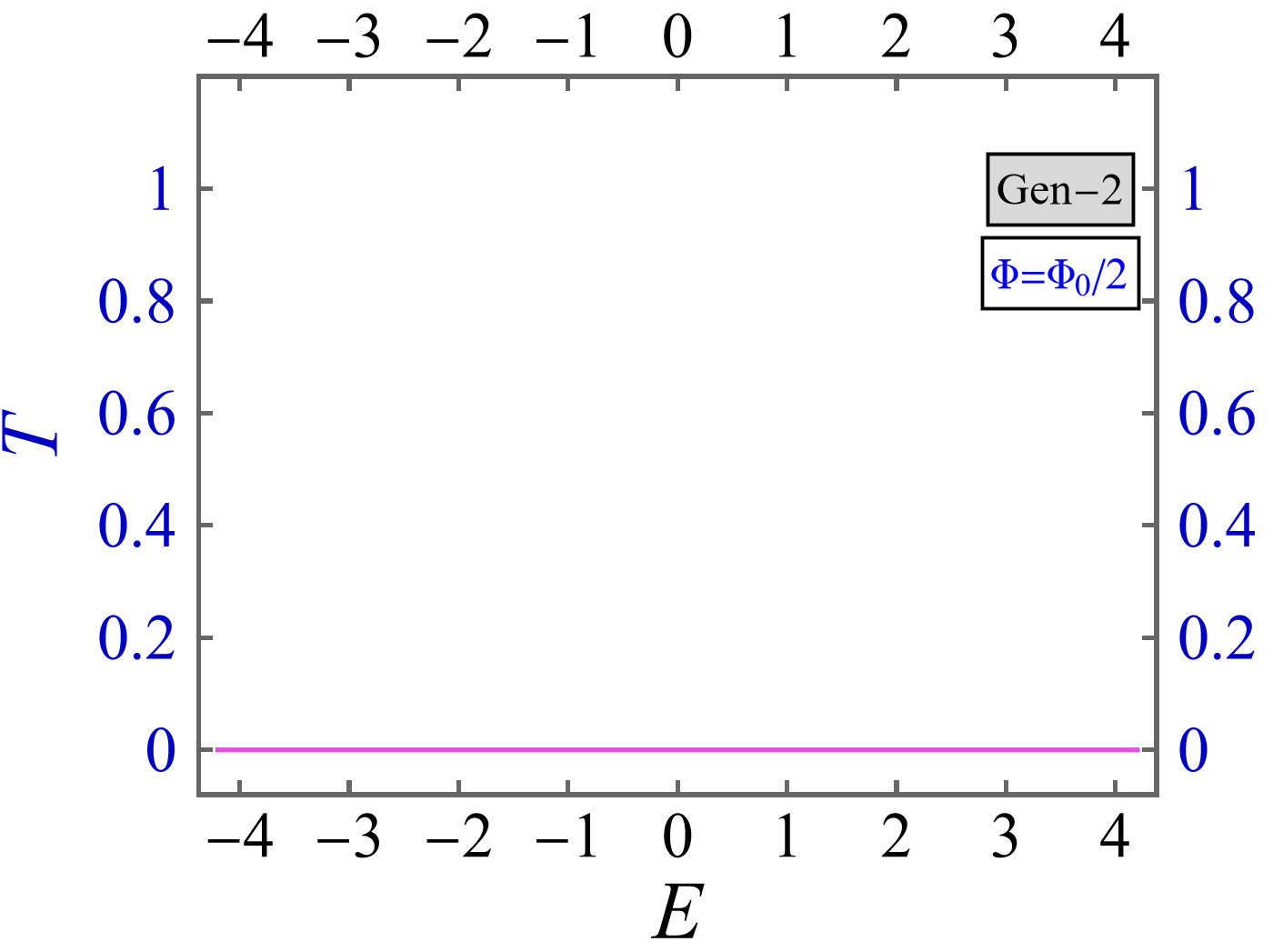}
\vskip 0.1cm
\textbf{(g)} \includegraphics[clip,width=0.6\columnwidth]{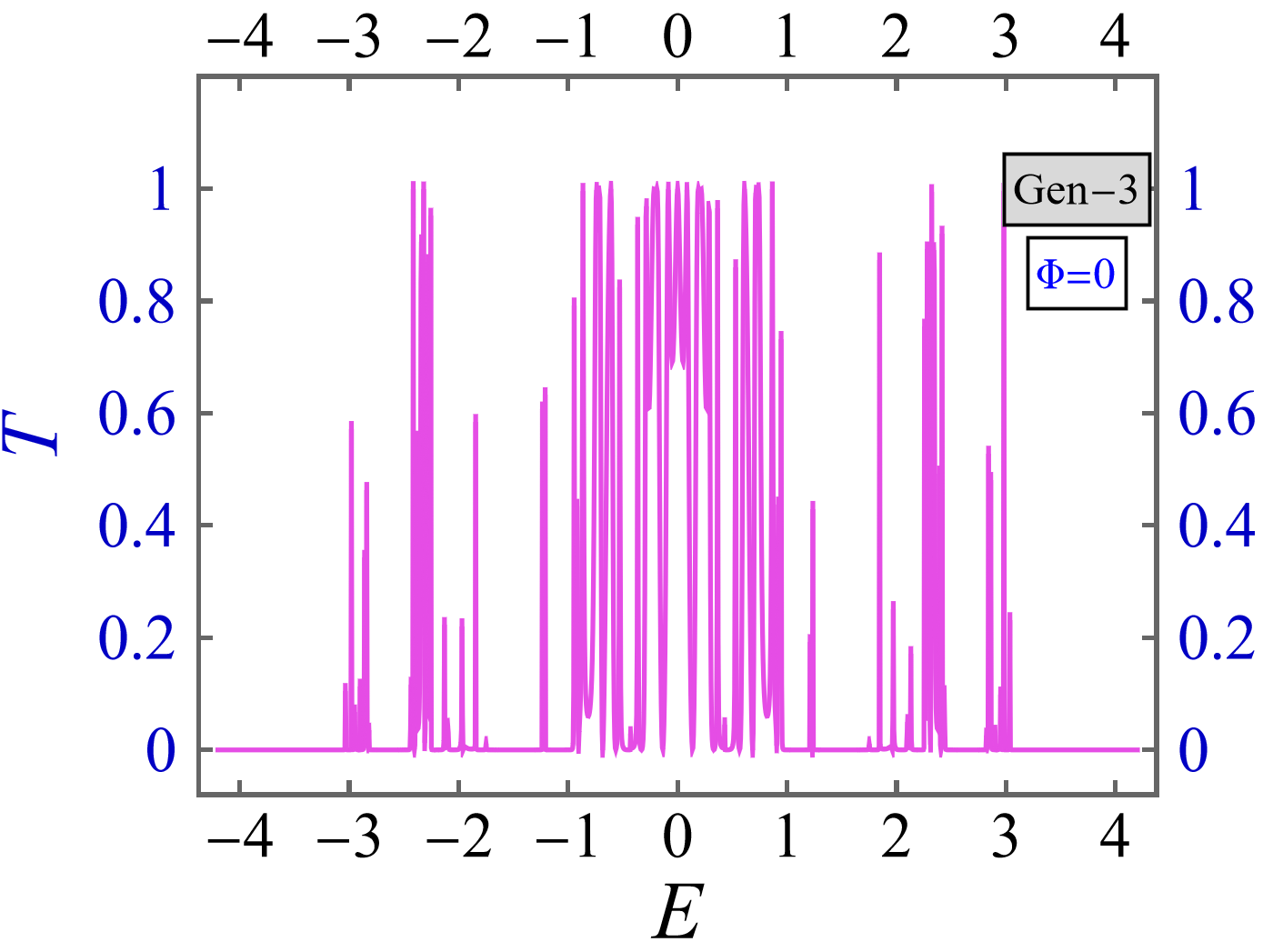}
\textbf{(h)} \includegraphics[clip,width=0.6\columnwidth]{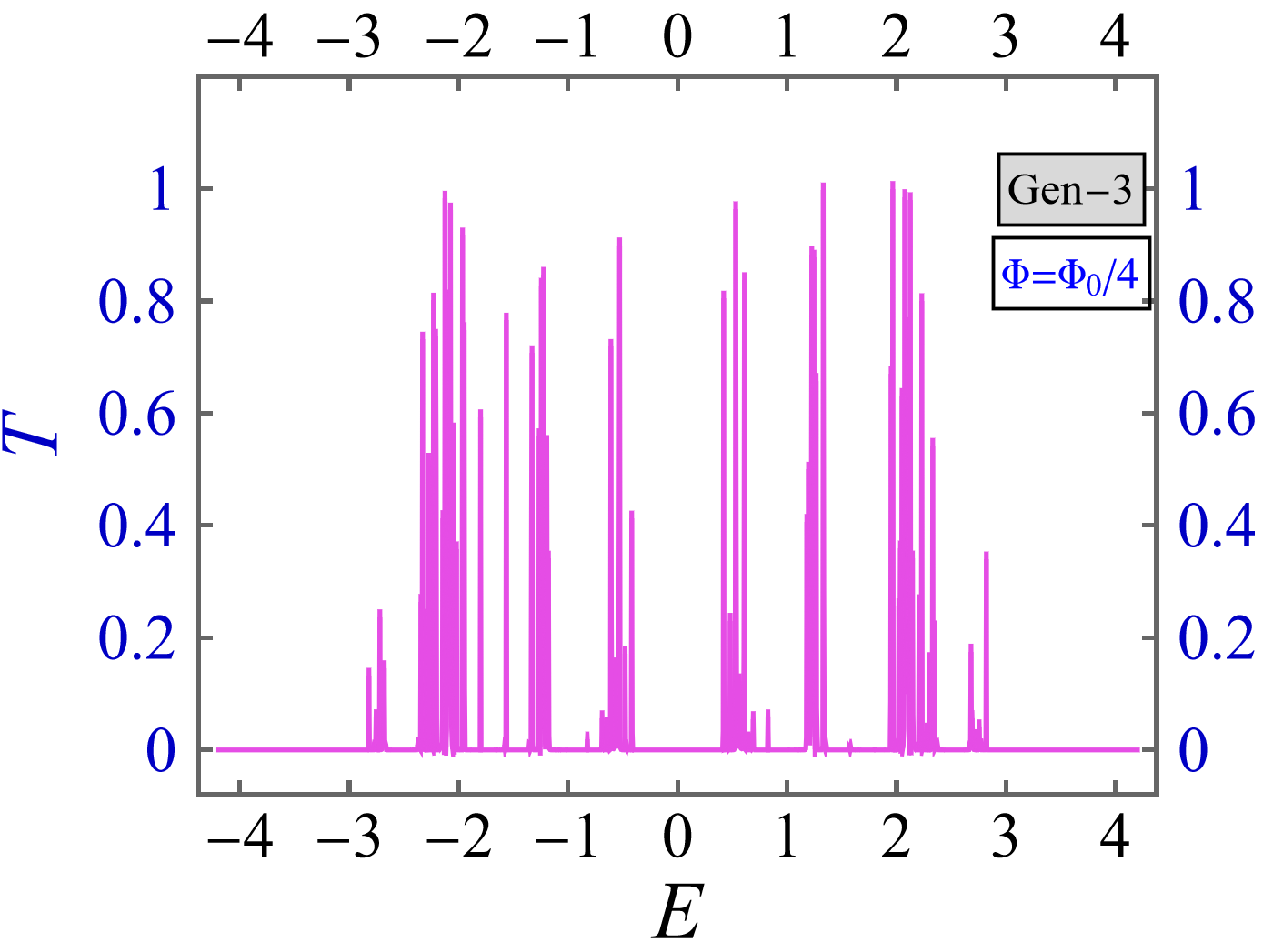}
\textbf{(i)} \includegraphics[clip,width=0.6\columnwidth]{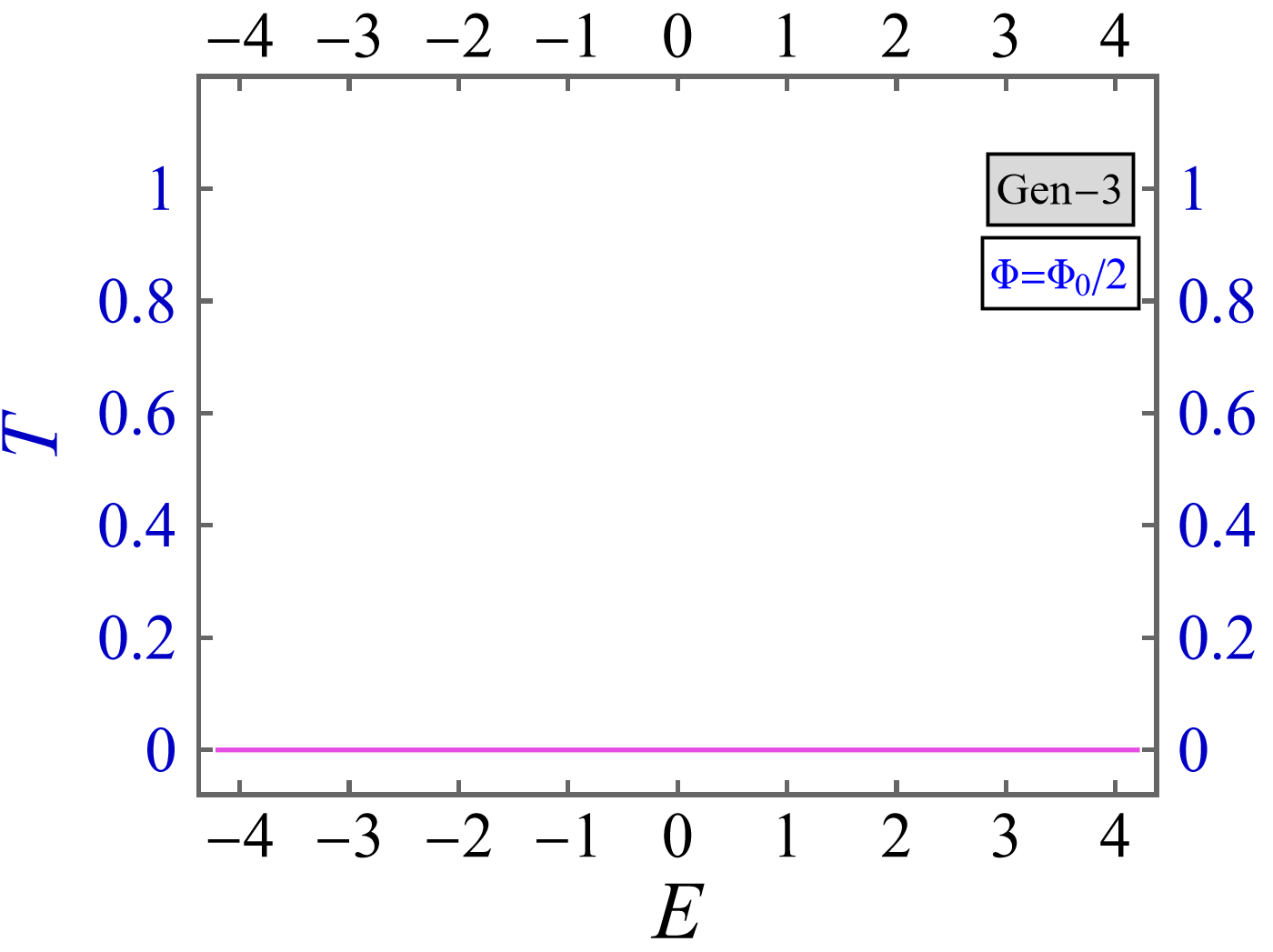}
\caption{The plots for the transmission probability ($T$) 
as a function of the energy ($E$) of the electron for the 
(a)-(c) first generation ($\mathcal{G}_{1}$), 
(d)-(f) second generation ($\mathcal{G}_{2}$), and 
(g)-(i) third generation ($\mathcal{G}_{3}$) Vicsek fractal lattice. 
The left column is for $\Phi=0$, 
the middle column is for $\Phi=\Phi_{0}/4$, and 
the right column is for $\Phi=\Phi_{0}/2$.}
\label{fig:transport}
\end{figure*}
%

In order to affirm our findings for the energy spectrum and the DOS, we 
have computed the two-terminal transmission probability of an electron 
through this fractal lattice system using the non-equilibrium Green's 
function (NEGF) technique~\cite{SDatta-book1,SDatta-book2,
Dey-OrgElectron-2011}. To find out the transmission probability of an 
electron through our fractal system, we connect a finite size fractal 
system in between two semi-infinite leads, which are popularly known as 
the source ($S$) and drain ($D$). We send the electron into the fractal 
system through the source and find the probability of it coming out of the 
system through the drain as we tune the value of magnetic flux in the system. 
The formula for the transmission probability is given by 
\begin{equation}
T(E) = \textrm{Tr}\big[ \bm{\Gamma_{S}G^{r} \Gamma_{D}G^{a}} \big],
\label{eq:Trans}
\end{equation} 
where $\bm{G^{r}}$ and $\bm{G^{a}}$ are the \textit{retarded} and 
\textit{advanced} Green’s function matrices for the system (including the 
coupling to the semi-infinite leads) given by
\begin{subequations}
\begin{align}
& \bm{G^{r}} = \big[ E{\bm I} - \bm{H} 
- \bm{\Sigma_{S}^{r}} - \bm{\Sigma_{D}^{r}} \big]^{-1} \\ 
& \bm{G^{a}} = \big[ E{\bm I} - \bm{H} 
- \bm{\Sigma_{S}^{a}} - \bm{\Sigma_{D}^{a}} \big]^{-1},
\label{eq:Greens-fn}
\end{align} 
\end{subequations}
and $\bm{\Gamma_{S}}$ and $\bm{\Gamma_{D}}$ are the broadening or coupling 
matrices representing the coupling between the system and the leads, 
and they are mathematically defined as
\begin{subequations}
\begin{align}
& \bm{\Gamma_{S}} = i \big( \bm{\Sigma_{S}^{r}}-\bm{\Sigma_{S}^{a}} \big) \\ 
& \bm{\Gamma_{D}} = i \big( \bm{\Sigma_{D}^{r}}-\bm{\Sigma_{D}^{a}} \big).
\label{eq:broadening-matrices}
\end{align} 
\end{subequations}
$\bm{\Sigma_{S(D)}^{r}}$ and $\bm{\Sigma_{S(D)}^{a}}$ are the retarded and 
advanced self-energies associated with the two leads, 
\textit{viz.}, the source and drain. 
The nonzero elements of the self-energy matrices are given by
\begin{subequations}
\begin{align}
& \bm{\Sigma_{S}^{r}}(p,p) = \dfrac{\tau_{S}^{2}}{2t_{L}^{2}}
\Bigg[\big(E-\varepsilon_{L}\big) 
- i \sqrt{4t_{L}^{2}-\big(E-\varepsilon_{L}\big)^{2}}\Bigg] \\ 
& \bm{\Sigma_{D}^{r}}(q,q) = \dfrac{\tau_{D}^{2}}{2t_{L}^{2}}
\Bigg[\big(E-\varepsilon_{L}\big) 
- i \sqrt{4t_{L}^{2}-\big(E-\varepsilon_{L}\big)^{2}}\Bigg] \\
& \bm{\Sigma_{S}^{a}}(p,p) = \big[\bm{\Sigma_{S}^{r}}(p,p)\big]^{\dagger} 
\ \textrm{\&} \ 
\bm{\Sigma_{D}^{a}}(q,q) = \big[\bm{\Sigma_{D}^{r}}(q,q)\big]^{\dagger} 
\label{eq:self-energy-matrices}
\end{align} 
\end{subequations}
where $p$ and $q$ denote the site numbers of the system with which 
the left lead (source) and right lead (drain) are connected, respectively. 
$\varepsilon_{L}$ is the constant onsite potential for the atomic sites in the 
leads, $t_{L}$ is the constant hopping integral between the neighboring sites 
in the leads, and $\tau_{S(D)}$ denote the coupling of the leads with the 
system. For our calculation, we set the parameters $\varepsilon_{L}=0$, 
$t_{L}=2$, and $\tau_{S(D)}=1$. 

Using the above prescription, we have calculated the transmission probability 
of an electron through the Vicsek fractal lattice for all the cases depicted 
in Fig.~\ref{fig:DOS}. We find that, for the cases 
$\Phi=0$ (Fig.~\ref{fig:transport} left column) and 
$\Phi=\Phi_{0}/4$ (Fig.~\ref{fig:transport} middle column), 
we get nonzero finite values of the transmission 
probability for an electron moving through the fractal system. 
The transmission probabilities show a lot of oscillations which is 
expected because of the fractal nature of the lattice system. However, our 
point of interest is $\Phi=\Phi_{0}/2$ (Fig.~\ref{fig:transport} right column). 
We find that, for the half flux quantum, \textit{i.e.}, $\Phi=\Phi_{0}/2$, we 
get zero transmission probability of the electron for the entire energy range. 
It means that, at $\Phi=\Phi_{0}/2$, all the single-particle states in the 
system are \emph{completely localized}. 
This quantifies and confirms our prediction of the AB caging phenomenon for 
this fractal geometry for a special value of the magnetic flux equal to 
half flux quantum. In Sec.~\ref{sec:persistent-current}, we discuss about the 
persistent current for the Vicsek fractal system, which in principle, could 
be measured in a real-life experiment.

\begin{figure}[ht]
\textbf{(a)} \includegraphics[clip,width=0.7\columnwidth]{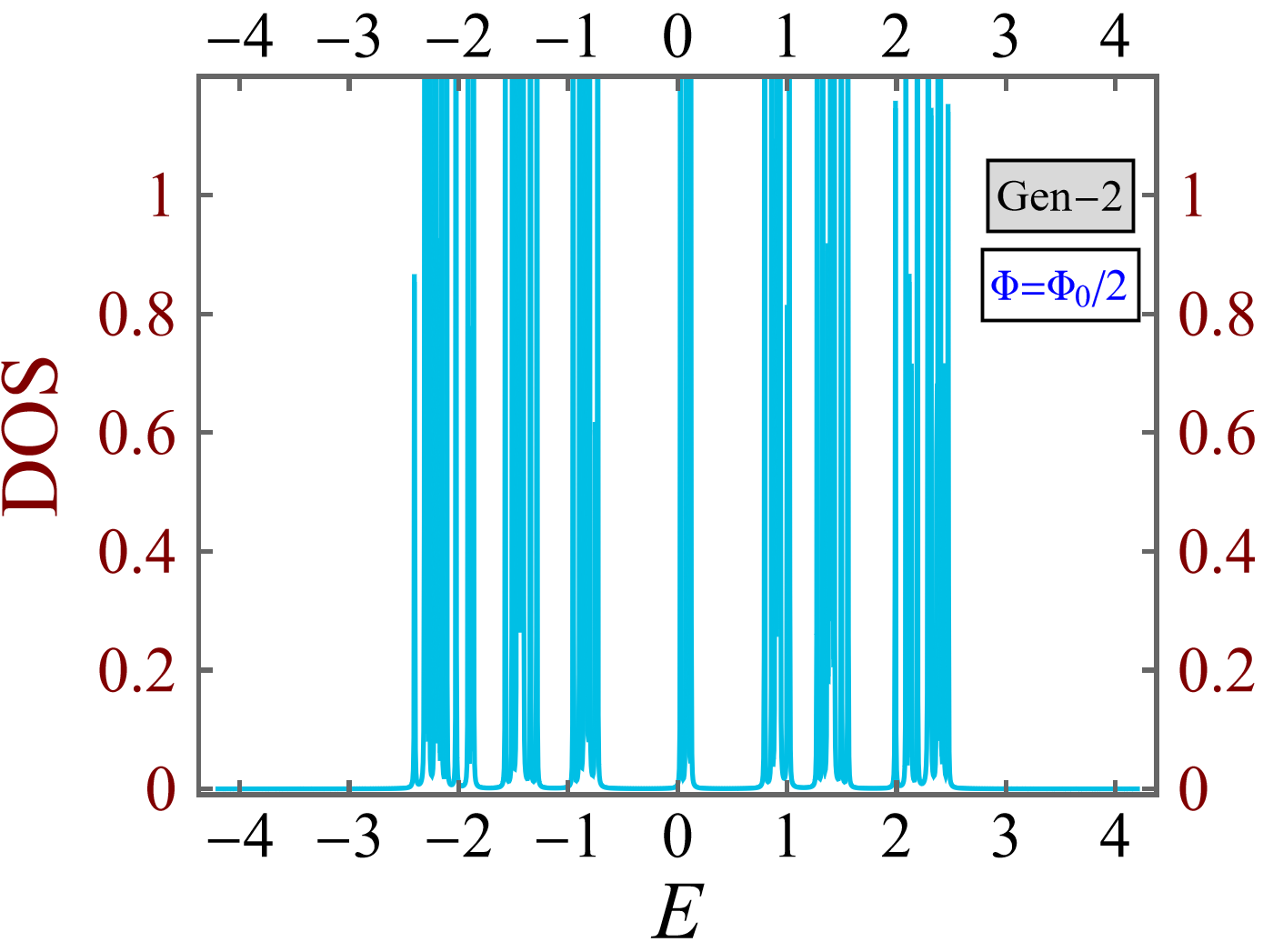}
\vskip 0.1cm
\textbf{(b)} \includegraphics[clip,width=0.7\columnwidth]{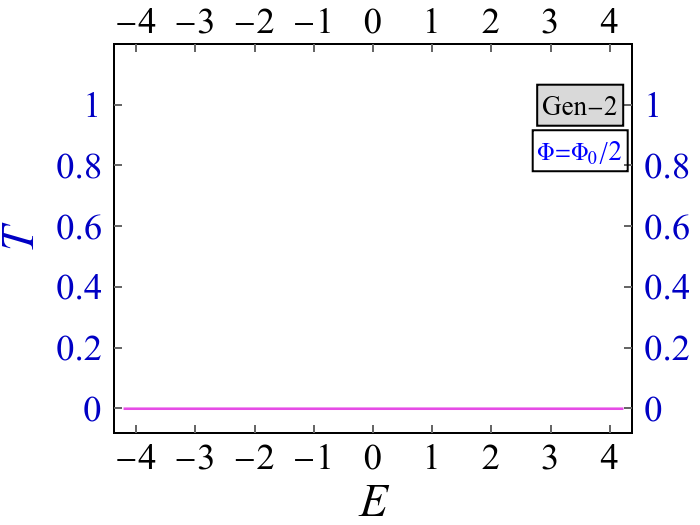}
\caption{The plots for the DOS and transmission probability ($T$) 
as a function of the energy ($E$) of the electron for the second 
generation Vicsek fractal lattice in presence of \emph{random onsite 
disorder} of width $\Delta=0.5$ in the system. 
We have chosen $\Phi=\Phi_{0}/2$.}
\label{fig:effect-of-disorder}
\end{figure}

\noindent
\textit{Effect of onsite disorder on the AB caging phenomena}: Before we end this 
section, let us briefly discuss the effect of the onsite disorder on the AB 
caging phenomena observed in a Vicsek fractal geometry. We will only focus on 
the case of half flux quantum, \textit{i.e.}, $\Phi=\Phi_{0}/2$. We have 
calculated the DOS and the two-terminal transport characteristics for this 
case with a random disorder in the onsite potential distribution 
$\varepsilon_{n} \sim \left[-\Delta/2,\Delta/2\right]$ of our system, 
where $\Delta$ is the width of the disorder strength. 
The results for the second generation 
fractal system are displayed in Fig.~\ref{fig:effect-of-disorder}. We find 
that there is a broadening in the DOS spectrum (see 
Fig.~\ref{fig:effect-of-disorder}(a)) which is resulted from the 
disorder distribution in the onsite potentials, however we still get 
zero transport (see Fig.~\ref{fig:effect-of-disorder}(b)) through the system 
even in presence of this random onsite disorder distribution.  
Thus, we can conclude that the phenomenon of AB caging of an electron in a 
Vicsek fractal lattice is robust against the onsite disorder in the system. 
It is observed that, at $\Phi=\Phi_{0}/2$ (half flux quantum), 
the effective hopping amplitude between the two horizontal sites of a 
single diamond plaquette becomes zero. That is why, even with the introduction 
of onsite disorder in the system, the AB caging phenomenon persist and we still 
get the zero transmission probability for the electrons in the system. 
One can expect to get the similar result even for the off-diagonal (hopping) 
disorder because of same logic as explained above.
%
%
\section{Persistent current}
\label{sec:persistent-current} 
%
\begin{figure}[ht]
\textbf{(a)} \includegraphics[clip,width=0.7\columnwidth]{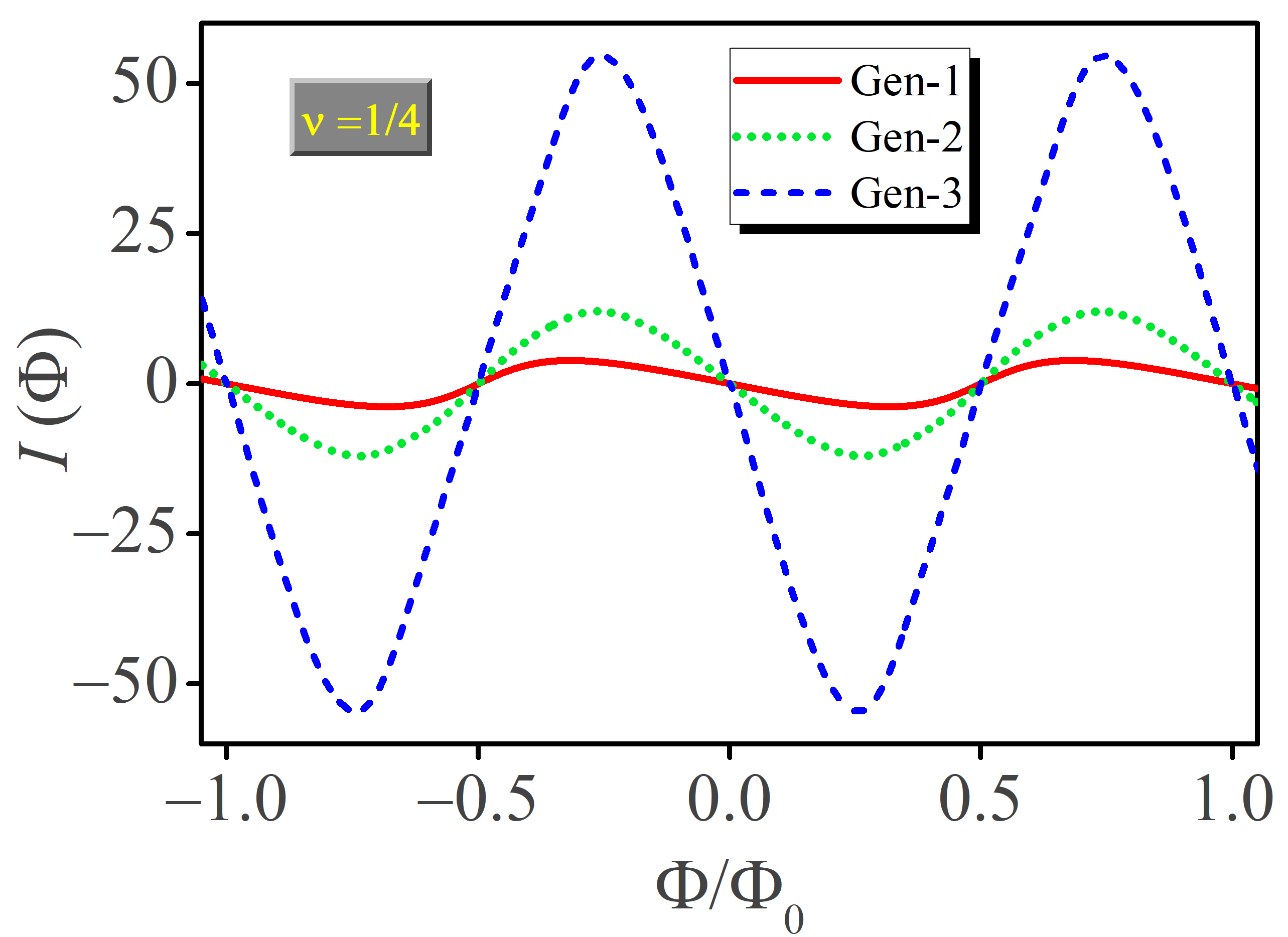}
\vskip 0.1cm
\textbf{(b)} \includegraphics[clip,width=0.7\columnwidth]{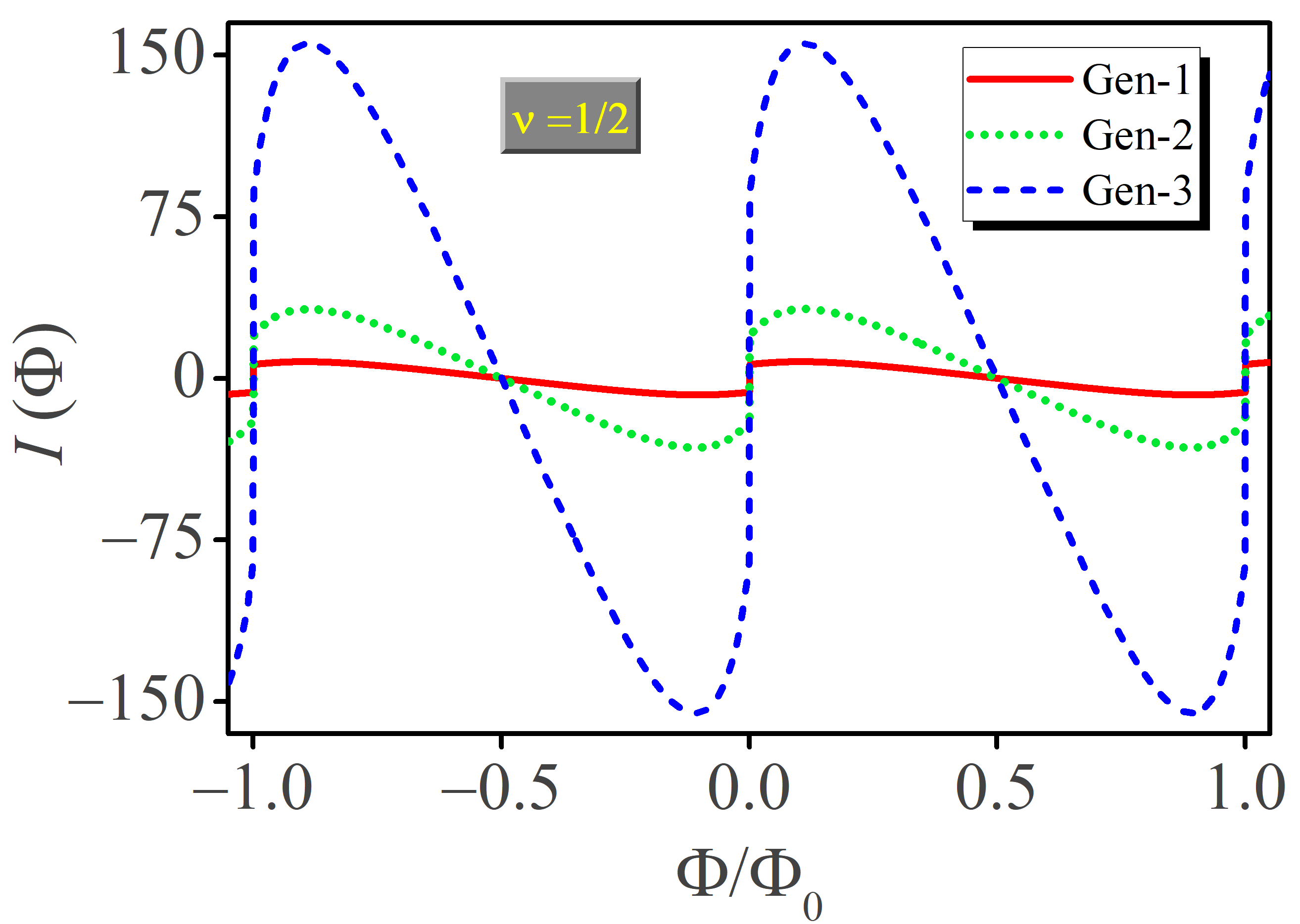}
\caption{Variation of the persistent current ($I$) as a function of the 
magnetic flux ($\Phi$) in a Vicsek fractal lattice for (a) $\nu=1/4$ filling 
and (b) $\nu=1/2$ filling of the energy levels.}
\label{fig:persistent-current}
\end{figure}
%
It is important to quantify the AB caging phenomena 
in terms of some experimentally measurable quantity, and the persistence current 
is a quantity which one can measure experimentally. Also, the study of the 
persistent current in this fractal system will give us a chance to explore the 
possibility of designing a quantum logic device using such a fractal system. 
Therefore, in this section, we discuss about the behavior of the 
persistent current in this closed-loop Vicsek fractal geometry. 
We remark that, it has no connection with the persistent current in the 
context of superconductivity. In our context, it is defined as a circulating 
current in a closed-loop geometry appearing because of the motion of an 
electron in the presence of a nonzero uniform magnetic flux in the system. 
The persistent current for the $n$th energy level in such 
closed-loop geometry at absolute zero temperature ($T=0$ K) is 
given by~\cite{Gefen-prl-1989,ac-prb-2010,biplab-jap-2025} 
\begin{equation} 
\mathcal{I}_{n} = -\dfrac{\partial E_{n}}{\partial \Phi}.
\label{eq:persistent-current}
\end{equation}
Hence, the total persistent current in the fractal system is given by, 
\begin{equation} 
I(\Phi) = \sum_{n=1}^{\mathcal{N}_{e}}\mathcal{I}_{n} = 
- \sum_{n=1}^{\mathcal{N}_{e}} \dfrac{\partial E_{n}}{\partial \Phi},
\label{eq:total-persistent-current}
\end{equation}
where $\mathcal{N}_{e}$ is the number of spinless non-interacting 
electrons in the system. 
We define $\nu = \mathcal{N}_{e}/\mathcal{N}_{\ell}$ as the 
\emph{filling factor} for an $\ell$th generation Vicsek fractal system. 

Using this formalism we have computed the persistent current in the 
Vicsek fractal lattice in its various generations for two different values 
of $\nu = 1/4$ and $1/2$ as exhibited in Fig.~\ref{fig:persistent-current}(a) 
and (b), respectively. We note that, in both the cases, the persistent 
current increases as the system size increases. In fact, from these two plots, 
we able to figure out a relationship between the persistent current in 
the $\ell$th generation Vicsek fractal and $(\ell+1)$th generation Vicsek 
fractal as follows: 
\begin{equation} 
I_{\ell+1} = R I_{\ell},\ \textrm{with}\ R=5\ \textrm{and}\ \ell\geq 1.
\label{eq:persistent-current-relationship}
\end{equation}
In Fig.~\ref{fig:persistent-current}, one can clearly identity that at 
$\Phi=\pm \Phi_{0}/2$, the persistent current vanishes in the system. 
Thus, this result corroborates with the phenomenon of AB caging for 
our fractal system and may be experimentally tested in the near future. 

Before we conclude, it is worth mentioning that similar 
lattice model has been studied earlier~\cite{pal-prb-2012} with a main focus 
to identify the staggered single-electron localization pattern in a Vicsek 
fractal geometry induced by the fractality of the lattice geometry in absence 
of any external magnetic flux. As compared to that, in the present study, our 
main focus is to understand the localization behavior, energy spectrum, 
transmission probability and persistent current in detail in a Vicsek 
fractal geometry as function of the magnetic flux. Secondly, in the earlier 
study, the real space renormalization group (RSRG) decimation method was used 
to obtain the results. In this method, some subtle features in the energy 
spectrum were missing because of the decimation of a certain set of sites. 
For example, in the present study, in Fig.~\ref{fig:eigenvalue-spectrum}, 
we find that a very robust flat band like energy eigenvalue appears in the 
energy spectrum at the energy $E=0$ starting from the second-generation 
system onwards. This interesting feature was not identified in 
Ref.~\cite{pal-prb-2012}. We have observed this in the present study, as 
here we have analyzed the energy spectrum using the exact diagonalization 
of the Hamiltonian matrices for different generation fractal lattice systems. 
Apart from that, in this work, we have quantified the AB caging phenomenon 
explicitly by calculating the transmission probability and the persistent 
current for various generations of the Vicsek fractal system. Additionally, 
we have also studied the effect of onsite disorder on the robustness of AB 
caging phenomena in such a fractal lattice.   
\section{Concluding remarks}
\label{sec:conclusion}
To conclude, we have demonstrated a mechanism by which one can completely 
block the mobility of an electron in a fractal geometry. Such caging of the 
electron happens for a special value of the underlying magnetic flux in the 
system and is governed by the geometrical structure of the basic building 
block of the corresponding fractal geometry. This intriguing phenomenon has 
been quantitatively verified by computation of the energy spectrum and the 
quantum transport for this fractal lattice model, and found to be 
robust against the onsite disorder in the system. Such controlled blocking 
of the quantum transport of an electron using an externally tunable 
magnetic flux could be very useful for the future quantum technology. 
For example, our results could be helpful in processing the quantum 
information in a quantum fractal system as well as for building 
efficient quantum algorithms using a quantum fractal network. In future, 
one can try to extend and generalize this idea of caging an electron in 
other possible closed-loop fractals.      
\begin{acknowledgments}
The author would like to thank Dr.\ Joydeep Majhi from IIT Bombay for 
some useful discussions during the course of this study.  
\end{acknowledgments}
%
\section*{Conflict of Interest}
The author declare no conflict of interest.
%
\section*{Data Availability Statement}
The data that support the findings of this study are
available from the corresponding author upon reasonable
request.

\end{document}